\documentclass[twocolumn,showpacs,amsmath,
amssymb,nofootinbib]{revtex4-1}
\usepackage{dcolumn}
\usepackage{bm}
\usepackage{graphicx}

\usepackage{color}

\newcommand{\be}{\begin{equation}}
\newcommand{\ee}{\end{equation}}
\newcommand{\ba}{\begin{eqnarray}}
\newcommand{\ea}{\end{eqnarray}}
\newcommand{\nn}{\nonumber\\}
\newcommand{\n}[1]{\label{#1}}
\newcommand{\eq}[1]{Eq.(\ref{#1})}

\newcommand{\hh}{\, ,\hspace{0.5cm}}
\newcommand{\hhh}{\, ,\hspace{0.2cm}}

\newcommand{\BM}[1]{{\mbox{\boldmath $#1$}}}

\newcommand{\bi}[1]{\bibitem{#1}}

\newcommand{\PRD}[4]{{#1}{Phys. Rev.\  D\ }{\bf #2},\  #3 (#4).}

\newcommand{\PRep}[4]{{#1}{Physics Reports\ }{\bf #2},\  #3 (#4).}

\newcommand{\APJ}[4]{{#1}{Astrophys. J.\ }{\bf #2},\  #3 (#4).}

\newcommand{\MN}[4]{{#1}{Monthly Notices of Royal Astr. Soc.\ }{\bf #2},\  #3 (#4).}

\newcommand{\NA}[4]{{#1}{Nature\ }{\bf #2},\  #3 (#4).}

\newcommand{\BOOK}[4]{{#1}{ {\it #2,\ }}{#3\ }{(#4)}}
\newcommand{\SC}[4]{{#1}{Science\ }{\bf #2},\  #3 (#4).}
\newcommand{\NJP}[4]{{#1}{New Journal of Physics\ }{\bf #2},\  #3 (#4).}
\newcommand{\RMP}[4]{{#1}{Reviews of Modern Physics\ }{\bf #2},\  #3 (#4).}
\newcommand{\ASA}[4]{{#1}{Astron. Astrophys.\ }{\bf #2},\  #3 (#4).}
\newcommand{\ASL}[4]{{#1}{Astronomy Letters\ }{\bf #2},\  #3 (#4).}
\newcommand{\AB}[4]{{#1}{Astrophysical Bulletin\ }{\bf #2},\  #3 (#4).}

\begin{document}

\title{Radiation from an emitter revolving around a magnetized non-rotating black hole}
\author{Valeri P. Frolov}
\email{vfrolov@ualberta.ca}
\author{Andrey A. Shoom}
\email{ashoom@ualberta.ca}
\author{Christos Tzounis}
\email{tzounis@ualberta.ca}
\affiliation{Theoretical Physics Institute, University of Alberta,
Edmonton, AB, Canada,  T6G 2E1}
\date{\today}

\begin{abstract}
One of the methods of study of black holes in astrophysics is based on broadening of the spectrum of radiation of  ionized Iron atoms.  The line K$\alpha$ associated with Iron emission at 6.4 keV is very narrow. If such an ion is revolving around a black hole, this line is effectively broadened as a result of the Doppler and gravitational redshift effects. The profile of the broaden spectrum contains information about the gravitational field of the black hole. In the presence of a regular magnetic field in the vicinity of a black holes the characteristics of the motion of charged ions are modified. In particular, their innermost stable circular orbits become closer to the horizon. The purpose of this work is to study how this effect modifies the spectrum broadening of lines emitted by such an ion.  Our final goal is to analyze whether the change of the spectrum profiles can give us information about the magnetic field in the black hole vicinity.
\end{abstract}

\pacs{04.20.Cv, 04.70.Bw, 04.70.-s, 04.25.-g \hfill
Alberta-Thy-09-14}

\maketitle

\section{Introduction}

There are more and more evidences that astrophysical black holes exist. Black hole candidates (both of stellar mass and supermassive) are identified by demonstration that a large mass compact object is located in a region of sufficiently small size, that practically excludes other different from a black hole models. Accretion of matter onto a black hole produces intensive radiation. By means of this radiation black holes manifest themselves. In particular such radiation may contain information about properties of the spacetime in the vicinity of a black hole and may confirm that a compact object is really a black hole. (For a comprehensive review of modern status of black holes in astrophysics see, e.g., a nice review \cite{Na:05}.)

One of the most useful methods is to use Iron K$\alpha$ lines as probes of the black hole vicinity (see, e.g., \cite{ReBe:77,BrChMi,ZaNuPaIn:05,Za:07} and references therein). This line of Iron has an energy that depends on the state of the ionization of the atom and is in the range of $6.40-6.97$ keV. Such a spectral line is excited as a result fluorescence of K$\alpha$ in a relatively cold accretion disc.
The line is intrinsically very narrow. When such an ion is revolving around a black hole, a distant observer registers that the line broadened as a result of the Doppler and gravitational redshift effects. It has a characteristic asymmetric double-peaked shape.
The form and the details of the broadened line depend on the parameters of the Keplerian orbit of the emitter.

There exist an important difference between Einstein and Newton gravity. Namely, for a particle moving in the gravitational field in the former case there exists the innermost stable circular orbit (ISCO). The radius of ISCO depends on the rotation parameter $a$. For a non-rotating (Schwarzschild, $a=0$) black hole of mass $M$ the ISCO radius is $6M$. In the presence of rotation (in the Kerr metric) it is closer to the black hole horizon, so that in the limit of the extremal rotation ($a=M$) ISCO radius becomes $M$. In principle, this allows one to use the method based on study of the spectral line broadening to determine a rotation parameter for a black hole. For a comparison with observations the broadened spectrum of Iron atoms emission should be additionally averaged over positions of the emitting atoms in the disk. This requires knowledge of the emissivity distribution on the disk (see, e.g., discussion in \cite{KaMi:04}).
The K$\alpha$ spectrum broadening from accretion disk around non-rotating and rotating black holes was calculated in \cite{LtoN95,CuBa:73,Cu:75,CaFaCa:98,FrKlNe:00,FuWu:04}.
A comprehensive review of the application of the fluorescent Iron lines as probe of properties of astrophysical black holes can be found in \cite{FaIwRe:00,ReNo:03,FuTs:04,Jo:12}.

In this paper we would like to discuss another interesting aspect of spectral line broadening method, namely usage of Iron K$\alpha$ lines as probes of the magnetic field in the black hole vicinity. There are (both theoretical and observational) evidences that magnetic field plays an important role in the black hole physics. Magnetic field seems to be essential to angular momentum transfer in accretion disks \cite{BaHa:98,Kr:01}.
Recent observations of the Faraday rotation of the radiation of a pulsar in the vicinity of a black hole in the center of the Milky Way (SgrA*) indicates that at a distance of few Schwarzschild radii  there exists magnetic field of several hundred Gauss \cite{Ea:13}. This supports emission models of SgrA* that requires similar magnetic field for explanation of the synchrotron radiation from a near horizon region (see, e.g., \cite{Fa,Mo,De}).

In the Blandford-Znajec model a regular magnetic field in the black hole vicinity is postulated to explain black hole jets energetics \cite{Zn:76,BlZn,MEM}. For example, in order to produce power of the magnitude $\sim 10^{45}$ erg/sec seen in the jets of supermassive (with mass $10^9 M_{\odot}$) rotating black holes,  regular magnetic field of the order $10^4$ G is required \cite{MEM}. Another mechanism for energy extraction from a rotating black hole and formation of relativistic jet, based on the analogue of the Penrose mechanism for a magnetic field was proposed in \cite{KoShKuMe:02,Ko:04}. The authors performed numerical simulations and demonstrated that the power in the jet emission as a result of such MHD Penrose process is of the same order as the estimate based
on the Blandford-Znajek mechanism. In particular, for a strong magnetic field $\sim 10^{15}$ G around a stellar-mass ($M\sim 10 M_{\odot}$) extremely rotating black hole the power of emission is estimated as $\sim 4\times 10^{52}$ erg/sec, which is similar to the power seen in gamma-ray bursts. Estimates based on the observed optical polarization for a number of active galactic nuclei gives the value $\sim10^3-10^5$ G for the magnetic field at the horizon of the corresponding black holes \cite{Si:09,Pi:00,SiGn:13}.

In this paper we discuss how a regular magnetic field in the vicinity of a black hole changes the parameters of the charged particle orbits. Our aim is to obtain the images of such orbits as they are seen by a far distant observer. We also study the  affect the broadening spectrum of  emission lines of Iron ions moving near nagnetized black holes. Let us mention that influence of a magnetic field on the distortion of the Iron K$\alpha$ line profile was earlier discussed in \cite{Za:03}. The authors focused on the  splitting of lines of emission due to Zeeman effect. They demonstrated that this effect might be important if the magnetic field is of the order $10^{10}-10^{11}$ G. We consider completely different effect, which might exist at much weaker magnetic fields. Namely, we assume that an Iron ion, emitting radiation, revolves around a magnetized black hole. For simplicity we assume that the black hole of mass $M$ is non-rotating. The Lorentz force, acting on a moving charged emitter in the magnetic field, modifies its motion.

The analysis of the corresponding equations of motion shows that the position of the ISCO  for charged particles is closer to the black hole horizon than ISCO radius for a neutral particle ($6M$) \cite{AG,FS}. This modification of the orbit is more profound in the case when the Lorentz force is repulsive. It is characterized by the dimensionless parameter
\be\n{bbb}
b={qBMG\over mc^4}.
\ee
Here $q$ and $m$ are charge and mass of the charged particle, $B$ is the strength of the magnetic field, and $M$ is the mass of the black hole. (Here we use CGS system of units.) The parameter $b$ is proportional to the ratio of the cyclotron frequency of a charged particle in the magnetic field $B$ in the absence of gravity, to the Keplerian frequency of a neutral particle at ISCO in the gravitational field of the black hole.
To estimate the value of this parameter one can consider a motion of a proton (mass $m_p=1.67\times 10^{-24}$g and charge $e=4.8\times 10^{-10} (\mbox{g\, cm}^3/\mbox{sec}^2)^{1/2}$). Then for a stellar mass black hole, $M=10 M_{\odot}$, this parameter takes  value $b=1$ for the magnetic field $B\sim 2$ G. For a supermassive black hole $M\sim 10^9 M_{\odot}$ the corresponding field is $B\sim 2\times 10^{-8}$ G. If the charge of the ion is $q=Ze$ and its mass is $m=Am_p$ the corresponding expression for the magnetic field parameter $b$ contains an additional factor $Z/A$. One can expect that for astrophysical black holes the parameter $b$ is large.

In fact, the magnetic field essentially modifies the orbits already when the parameter $b$ is of the order of 1. For the repulsive Lorentz force case the radius of ISCO in the strong magnetic field ($b\gg1$) can be located arbitrary close to the horizon. As a result, two new effects are present in the process of emission of an ion of Iron in magnetized black holes: (i) position of ISCO depends on the magnetic field, and (ii) even for a circular motion of the same radius, angular velocity of charged particles differs from the Keplerian angular velocity. In this paper we study how these effects modify the broadening of the K$\alpha$ spectrum.

The paper is organized as follows. In Sec.~II we remind the main formulas concerning the charged particle motion near a magnetized black hole. In Sec.~III we collect results for ray tracing in the Schwarzschild geometry, which are required for construction of the orbit's images and the calculations of the spectrum broadening.
In Section~IV we discuss the particle orbits images. In Sec.~V. we derive an expression for a broadened spectrum for the sharp spectral lines emitted by  the point-like source moving at the circular orbit near a magnetized black hole.
The numerical results for the spectral function are presented in Sec.~VI and discussed  in Sec.~VII. The latter section contains also general discussion of the results of the present paper and their possible extensions.
Some auxiliary  calculations are collected in appendices.

In this paper we use the sign conventions adopted in \cite{MTW} and units where $G=c=\hbar=1$.

\section{Charged particles in magnetized black holes}

\subsection{Magnetized black hole}

We consider a magnetized non-rotating black hole. We assume that magnetic field is weak, so that its back reaction on the black hole's spacetime geometry can be neglected. The metric of the black hole is
\be
dS^2=-f dT^2+{dr^2\over f}+r^2 d\Omega^2\hh f=1-{r_g\over r}\, ,\n{1}
\ee
where $r_g=2M$ is the black hole's gravitational radius, and
\be\n{1d}
d\Omega^2=d\theta^2+\sin^2\theta d\phi^2\,
\ee
is the metric on a unit sphere $S^2$.
Such a spacetime has only one dimensional parameter, $r_g$, and one can write the metric in the form
\be
dS^2=r_g^2 ds^2\hh
ds^2=-f dt^2+{d\rho^2\over f}+\rho^2 d\Omega^2\, ,
\ee
where $t=T/r_g$ and $\rho=r/r_g$ are dimensionless time and radius and $f=1-1/\rho$. In what follows, we shall use this dimensionless form of the metric.

The metric $ds^2$ possesses four Killing vectors
\ba
&&\BM{\xi}_{(t)}=\partial_t\hh \BM{\xi}_{(\phi)}=\partial_{\phi}\, ,\n{sym1}\\
&&\BM{\xi}_{x}=  -\cos\phi \partial_{\theta}+\cot\theta \sin\phi \partial_{\phi}\, ,\n{sym2}\\
&&\BM{\xi}_{y}= \sin\phi \partial_{\theta}+\cot\theta \cos\phi \partial_{\phi}\, .\n{sym3}
\ea
The first one is the generator of time translations, while the other three are the generators of rotations.

We choose the magnetic field of the form
\be
A^{\mu}=\frac{B r_g}{2}\xi^{\mu}_{(\phi)}\, ,\n{2}
\ee
where $B=const$ (see, e.g., \cite{AG,Wald}).
It is static and axisymmetric, and it is homogeneous at the asymptotic infinity $(\rho\to+\infty)$ with the strength $B$ and directed orthogonal to the equatorial plane $\theta=\pi/2$.
The electromagnetic field tensor $F_{\mu\nu}$ has the following form:
\ba
F_{\mu\nu}&=&2A_{[\nu,\mu]}\nonumber\\
&=&2Br\sin\theta\left(\sin\theta\delta^{r}_{[\mu}\delta^{\phi}_{\nu]}+r\cos\theta
\delta^{\theta}_{[\mu}\delta^{\phi}_{\nu]}
\right)\,.\n{4}
\ea

\subsection{Equatorial motion of a charged particle}

In this paper, we shall adopt a number of simplifying assumptions. We consider a freely moving single Iron ion, which emits a sharp spectral line, and we neglect its interaction with other matter, surrounding a black hole.
Then a charged particle motion obeys the equation
\be
{D u^{\mu}\over d\tau}={q\over m} F^{\mu}_{\ \nu} u^{\nu}\, .\n{3}
\ee
Here $u^{\mu}=dx^{\mu}/d\tau$ is the particle 4-velocity, $u^{\mu}u_{\mu}=-1$, $\tau$ is its dimensionless proper time, $q$ and $m$ are its electric charge and mass, respectively.

For the motion around the magnetized black hole there
exist two conserved quantities associated with the Killing vectors \eq{sym1}:
the specific energy ${\cal E}>0$ and the specific generalized azimuthal angular momentum
$l\in(-\infty,+\infty)$,
\ba
&&{\cal E}\equiv -\xi^{\mu}_{(t)}u_{\mu}={d t\over d\tau}\left(1-\frac{1}{\rho}\right)\,,\n{5}\\
&&l\equiv \xi^{\mu}_{(\phi)}(u_{\mu}+{q\over m}A_{\mu})=\left({d \phi\over d\tau}
+b\right)\rho^2\sin^2\theta\,.\n{6}
\ea
Here the parameter $b$ characterizes the dimensionless strength of the magnetic field [see \eq{bbb}].

It is easy to check that the $\theta$-component of \eq{3} allows
for a solution $\theta=\pi/2$. This is a motion in the equatorial plane of the black hole, which is orthogonal to the magnetic field. Here we restrict ourselves to this type of motion for which the conserved quantities \eq{5} and \eq{6} are sufficient for the complete integrability of the dynamical equations.

The complete integrability allows one to write the equations of motion  in the equatorial plane in the following first order form
\ba
&&\hspace{0.8cm}\left(\frac{d\rho}{d\tau}\right)^2={\cal E}^2-U\,,\n{12}\\
&&\frac{d\phi}{d\tau}=\frac{l}{\rho^2}-b
\hh {dt\over d\tau}=\frac{{\cal E}\rho}{\rho-1}\,.\n{13}
\ea
Here
\be\n{14}
U=\left(1-\frac{1}{\rho}\right)
\left[1+\frac{(l-b\rho^2)^2}{\rho^2}\right]\,.
\ee
is the effective potential.

For the motion in the equatorial plane the equations
are invariant under the following transformations:
\be\n{10}
b\to- b\hhh l\to-l\hhh \phi\to -\phi\, .
\ee
Thus, without loss of the generality, one can assume that the charge
$q$ (and hence $b$) is positive. For a particle with a negative charge it is sufficient to make the transformation \eq{10}. According to the adopted convention, we
have $b\ge 0$. The parameter $l$ can be positive or negative.
For $l>0$ (sign $+$) the Lorentz force, acting on a charged
particle, is repulsive, i.e., it is directed outward from the black
hole. Following the paper \cite{AG}, we call such motion {\em anti-Larmor motion}. In the opposite case when $l<0$ (sign $-$) the Lorentz force is attractive,
i.e., it is directed toward the black hole. We call it {\em Larmor motion}.

\subsection{Stable circular orbits (SCO's)}

The equatorial motion around the magnetized Schwarzschild black hole \eq{1} was studied in detail in the papers \cite{FS,FF}. We remind here some properties of the circular motion, following this paper. For the circular motion the radial coordinate is fixed $\rho=\rho_{e}=const$.

Extrema of the effective potential are defined by the equation $U_{,\rho}=0$.  We assume that $U_{,\rho\rho}\ge 0$, so that this is a local minimum and the corresponding circular orbit is stable. We call such an orbit a {\em stable circular orbit} or briefly SCO  The equation $U_{,\rho}=0$ allows us to find the parameter $l$ as a function of $\rho_{e}$ and $b$,
\be\n{lr}
l_{\pm}=\frac{-b\rho_{e}^{2}\pm \rho_e\sqrt{2\rho_e-3+4b^2\rho_e^2(\rho_e-1)^2}}{(2\rho_{e}-3)}\, .
\ee
Accordingly, the specific energy of the particle at SCO is
\be\n{er}
{\cal E}=\sqrt{U(\rho_{e})}=\left(1-\frac{1}{\rho_{e}}\right)^{1/2}
\left[1+\frac{(l_{\pm}-b\rho_{e}^2)^2}{\rho_{e}^2}\right]^{1/2}\,.
\ee

Equations \eq{13} determine the angular position of the particle $\phi=\Omega t$.  We choose the coordinate $\phi$ so that the motion with $l>0$ is counterclockwise and $\varphi$ changes in the interval $(-\pi,\pi]$.
The corresponding dimensionless angular frequency of the particle motion is
\be\n{11}
\Omega=\frac{d\varphi}{dt}=\frac{\rho_{e}-1}{\rho_{e}{\cal E}}\left(\frac{l}{\rho_{e}^2}-b\right)\,.
\ee
Substituting  expressions \eq{lr} and \eq{er} into \eq{11} we find the angular frequency $\Omega$ as a function of $\rho_{e}$ and $b$. In this expression, for a fixed value of the parameter $b\geq0$, the specific energy ${\cal E}$ and the parameter $l$ are defined by the value of $\rho_{e}$, which corresponds to minimum of the effective potential.

For $\rho_e$ greater than the ISCO radius this is a local minimum. If the specific energy is greater than the value of the potential at this minimum, the radial motion is an oscillation between the minimal and maximal values of the radius. As a result the motion with negative $l$ remains smooth, while for $l>0$ and large enough magnetic field $b$ the particle trajectory becomes curly. One can describe such a trajectory as a result of superposition of cyclotron rotation along small cycles and a slow drift motion of the center of the cycle around the black hole. One can expect that as a result of the synchrotron radiation such a trajectory would become more smooth and finally relax  to a circular one. For more details concerning general type of motion in magnetized black holes, see \cite{FS}.

\subsection{Innermost stable circular orbits (ISCO's)}

For a given magnetic field $b$ there exist a minimal radius of SCO. For circular orbits with smaller radius the motion becomes unstable. The corresponding innermost stable circular orbit is known as ISCO. The ISCO radius is a simultaneous solution of the following two equations
\be
U_{,\rho}=0\hh U_{,\rho\rho}=0\,.
\ee

\begin{figure}[htb]
  \includegraphics[width=5cm]{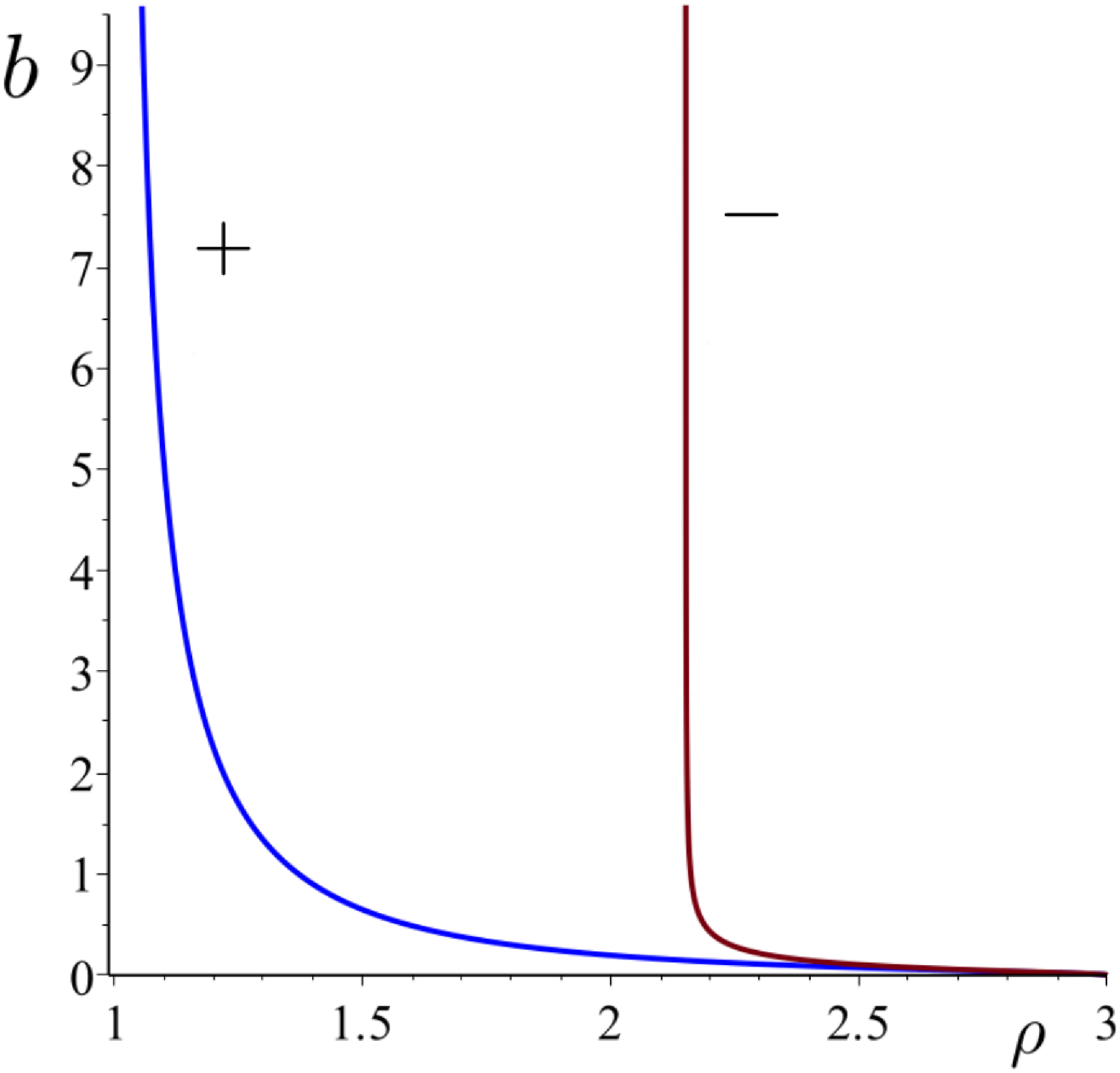}\\
  \caption{Magnetic field $b$ of a magnetized black hole is a function of ISCO radius $\rho$. Labels $+$ and $-$ stand for the anti-Larmor and Larmor orbit branches, respectively.}\label{brho}
\end{figure}

In this case, the parameter $b$ is not free and it depends on the value of $\rho_{\pm}$. As a result, for ISCO we have,
\ba
l_{\pm}&=&\pm\frac{\rho_{\pm}(3\rho_{\pm}-1)^{1/2}}{\sqrt{2}
H_{\pm}}\, ,\ 
b=\frac{(3-\rho_{\pm})^{1/2}}{\sqrt{2}\rho_{\pm}H_{\pm}}\,, \n{pm1}\\
\Omega_{\pm}&=& \pm \frac{\sqrt{2}}{2}\frac{\sqrt{3\rho_{\pm}-1}\mp \sqrt{3-\rho_{\pm}}}{J_{\pm}} \, , \\
 {\cal E}_{\pm} &=& \sqrt{\frac{\rho_{\pm}-1}{\rho_{\pm}}}\frac{J_{\pm}}{H_{\pm}}  \, .\n{pm2}
\ea
Here
\ba
J_{\pm}&=&\sqrt{H_{\pm}^{2}+\rho_{\pm}+1\mp\sqrt{3\rho_{\pm}-1}\sqrt{3-\rho_{\pm}}} \, ,\\
H_{\pm}&=&\sqrt{4\rho_{\pm}^2
-9\rho_{\pm}+3\pm\sqrt{(3\rho_{\pm}-1)(3-\rho_{\pm})}} ,
\ea
and $(5+\sqrt{13})/4<\rho_{-}\leq3$ and $1<\rho_{+}\leq3$.

\begin{figure}[htb]
  \includegraphics[width=5cm]{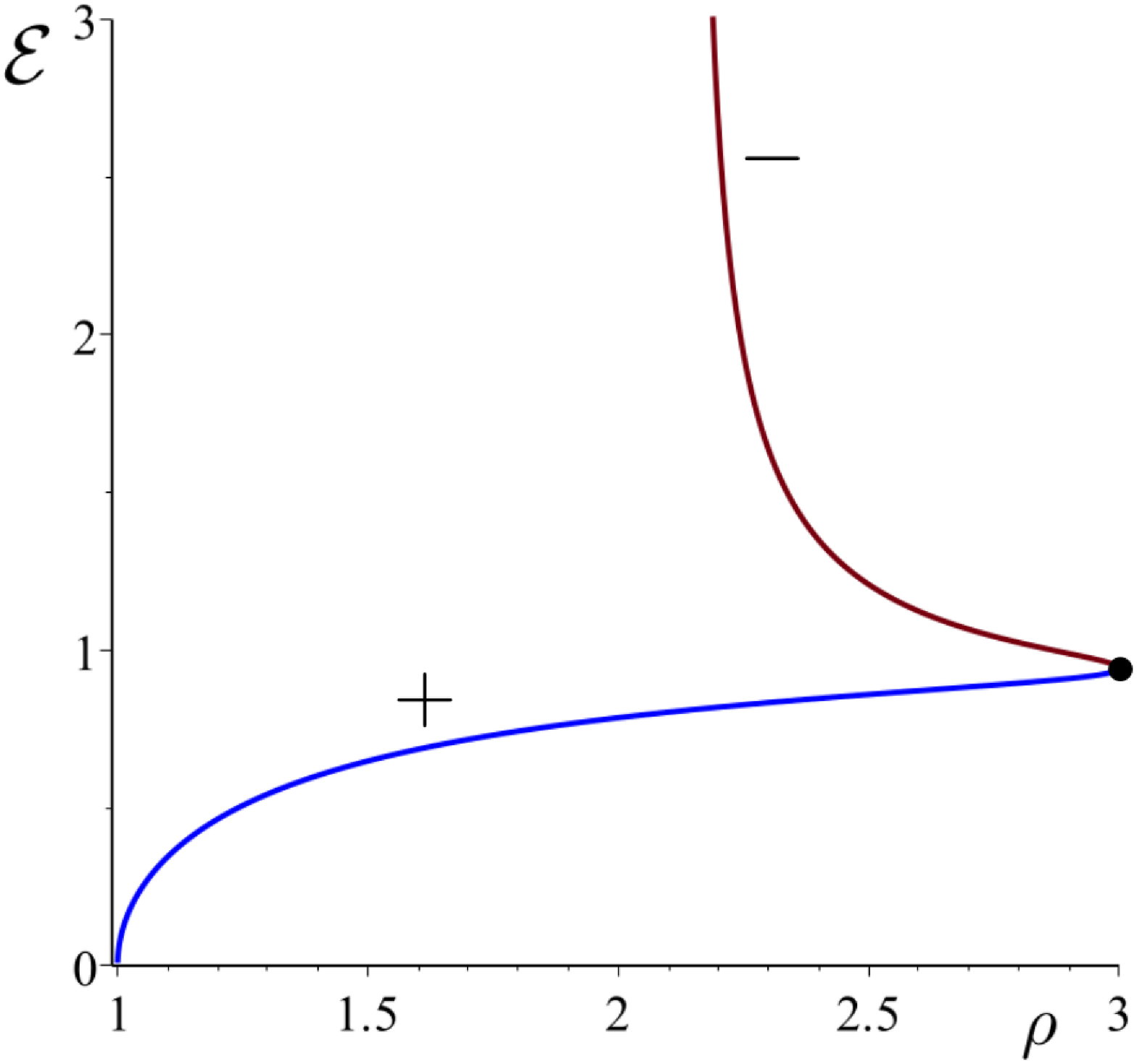}\\
  \caption{Specific energy ${\cal E}$ of a charged particle  at ISCO  in a magnetized black hole as a function of ISCO radius $\rho$. Labels $+$ and $-$ stand for the anti-Larmor and Larmor orbit branches, respectively.}\label{Energy}
\end{figure}

\begin{figure}
  \includegraphics[width=5cm]{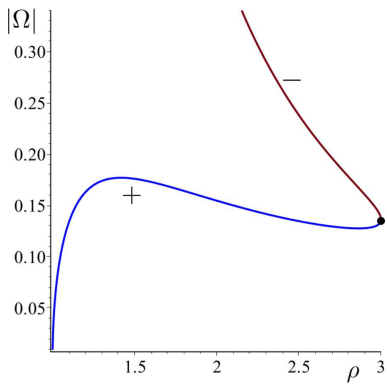}\\
  \caption{Angular velocity $|\Omega|$ as a function ISCO radius in a magnetized black hole. Labels $+$ and $-$ stand for the anti-Larmor and Larmor orbit branches, respectively.}\label{Omega}
\end{figure}

Figure~\ref{brho} shows a relation between the value of the magnetic field and the radius $\rho$ of the corresponding ISCO. Labels $+$ and $-$ stand for anti-Larmor and Larmor orbits, respectively. Specific energy at ISCO orbits as a function of the ISCO radius for both types (anti-Larmor and Larmor) of motion is presented in Figure~\ref{Energy}. The next plot (Figure~\ref{Omega}) shows the angular velocity at ISCO $|\Omega|$ as a function of its radius $\rho$. For small $b$ ($\rho\approx 3$) both branches $+$ and $-$ approach the same value $\Omega_{\mbox{ISCO},b=0}=1/\sqrt{54}\approx 0.136$, which is the Keplerian ISCO angular velocity for a non-magnetized black hole.

The asymptotics of these functions for anti-Larmor ISCO and  large $b$ are\footnote{In this paper, we focus on the anti-Larmor orbits which can be arbitrary close to the horizon of the black hole. However, similar expressions can be easily obtained in the large $b$ limit for Larmor orbits. For example, one has
\ba
&&\Omega_{-}|_{b> > 1}= \Omega_{-}^{0} +  \frac{\Omega_{-}^{2}}{b^2}+\mathcal{O}(b^{-4})\, ,\\
&&\Omega_{-}^{\left(0\right)}=-\frac{\sqrt{6}}{18}\sqrt{3+\sqrt{13}}
\left(19-5\sqrt{13}\right)\approx -0.34\, ,\\
&&\Omega_{-}^{\left(2\right)}=\frac{\sqrt{6}}{108}\sqrt{\frac{3+\sqrt{13}}{13}}
\left(7\sqrt{13}-25\right)\approx 0.41\, .
\ea
}
\ba
&&(\rho_{+}-1)|_{b>> 1}={1\over \sqrt{3}b}+\ldots\, ,\n{bbbb}\\
&&\Omega_{+}|_{b>> 1}=\frac{3^{3/4}}{6\sqrt{b}}+\ldots \ ,\\
&&{\cal E}_{+}|_{b>> 1}=\frac{2}{3^{3/4}\sqrt{b}}-\frac{2}{3^{5/4}b^{3/2}}+\ldots \, .
\ea
In the limit of the strong magnetic field ($b\gg 1$) $\Omega_+ \to 0$ (branch $+$) and $-\Omega_- \to 0.34$ (branch $-$).

\subsection{Circular motion in the rest frame basis}

Let us introduce a local rest frame $\{\BM{e}_t,\BM{e}_{\rho},\BM{e}_{\theta},\BM{e}_{\phi}\}$
\ba
\BM{e}_t&=&|\BM{\xi}_t^2|^{-1/2} \BM{\xi}_t=f^{-1/2}\partial_t\hh
\BM{e}_{\rho}=f^{1/2}\partial_{\rho}\, ,\\
\BM{e}_{\theta}&=&\rho^{-1}\partial_{\theta}\hh
\BM{e}_{\phi}=|\BM{\xi}_{\phi}^2|^{-1/2} \BM{\xi}_{\phi}={1\over \rho \sin\theta}\partial_{\phi}\, .
\ea

\begin{figure}
  \includegraphics[width=5cm]{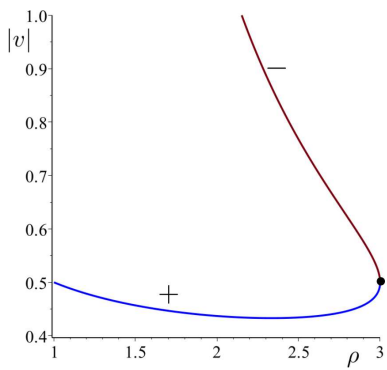}\\
  \caption{Velocity $v$ of a charged particle  at the ISCO  in a magnetized black hole at a function of ISCO radius $\rho$. Labels $+$ and $-$ stand for the anti-Larmor and Larmor orbit branches, respectively.}\label{Velocity}
\end{figure}

The four-vector of velocity for the circular motion with (dimensionless) angular velocity $\Omega$ can be written as follows
\ba\n{f1}
&&u^{\mu}={\gamma}\left(\xi^{\mu}_{(t)}+\Omega \xi^{\mu}_{(\phi)}\right)
= \tilde{\gamma} (e_{(t)}^{\mu}+v e_{(\phi)}^{\mu})\,,\\
&&{\gamma} =\frac{1}{\sqrt{f-\Omega^{2}\rho^{2}}}\hhh v={\Omega \rho\over \sqrt{f}}\hhh
\tilde{\gamma}=\frac{\sqrt{f}}{\sqrt{f-\Omega^2\rho^2}} \, .\n{pp}
\ea
Here $v$ (which can be either positive or negative) is the velocity of the particle with respect to a rest frame, and
$\tilde{\gamma}=(1-v^2)^{-1/2}$ is the corresponding Lorentz gamma factor.
A simple analysis shows, that  the velocity $v$ at the anti-Larmor ISCO remains close to $1/2$ in the entire interval $(1,3)$ of ISCO radii and, hence, this motion is not very relativistic. For the opposite direction of motion, the velocity for ISCO changes from $1/2$, in the absence of the magnetic field, up to 1 for very large magnetic field (see Figure~\ref{Velocity}). For more details see \cite{FF}.

\subsection{Near-horizon orbits}

Denote by $z$ a proper distance from the black hole's horizon
\be
z=\int_{r_g}^r {dr\over \sqrt{1-r_g/r}}\, .
\ee
At the horizon $z=0$ and in its vicinity one has
\be
r-r_g\sim {1\over 2}\kappa z^2\, ,
\ee
where $\kappa=1/(2r_g)$ is the surface gravity of the black hole.
We consider now a limit when $z$ is fixed, while $z/r_g\to 0$. We consider a space region near the equatorial plane which has the size in the orthogonal to $z$ direction much less than $r_g$. Denote  $d\eta=\kappa dT$, $dx=r_g d\theta$ and $dy=r_g d\phi$, then the metric \eq{1} in such a domain can be written as follows
\be
dS^2=-z^2 d{\eta}^2+dz^2+dx^2+dy^2\, .
\ee
This is the Rindler metric. In the spacetime domain covered by the Rindler coordinates the ISCO is described by a simple equations: $x=0$, $dy/d\eta =$\,const, $z=$\,const. Such a particle moves in the $y-$direction with a constant velocity $v$ [see \eq{f1}]. The equation of the charged particle motion \eq{3} implies
\be\n{eqq}
\Gamma_{z,\eta\eta}\dot{\eta}^2={q\over m} F_{z y}\dot{y}\, .
\ee
This equation has a simple meaning: In order  to stay at  $z=$\,const, the gravitational attraction force [the left-hand side of \eq{eqq}] must be compensated by the repulsive Lorentz force [the right-hand side of \eq{eqq}]. Simple calculations give
\ba\n{eeq}
\Gamma_{z,\eta\eta}&=&z\hh F_{zy}={z\over r_g^2}F_{r\phi}={Bz\over r_g}\, ,\\
\dot{y}&=&{v\over \sqrt{1-v^2}}\hh \dot{\eta}={\sqrt{1-v^2}\over z}\, .\n{eeq1}
\ea
The magnetic field in the Rindler domain $\vec{B}$ has only one component
\be
B^x=-{Bz\over r_g}\, ,
\ee
so that the Lorentz force acting on a particle moving in the $y-$direction with the velocity $\vec{v}$ is
\be
\vec{F}=q \vec{v}\times \vec{B}\, .
\ee
As expected, it is directed along the $z-$axis, that is away from the horizon.
After substitution of the relations (\ref{eeq})  and (\ref{eeq1}) into \eq{eqq}] one obtains
\be
bv\sqrt{1-v^2}\left(\frac{r}{r_{g}}-1\right)=1\, .
\ee
Since close to the horizon $v=1/2$, one gets
\be
\rho-1={1\over \sqrt{3}b}\, .
\ee
This formula correctly reproduces the asymptotic relation (\ref{bbbb}).

\subsection{Main properties of anti-Larmor particle motion in magnetized black holes}

Let us summarize. The ISCO radius for charged particles in the magnetized black holes is smaller than $6M$ in both the cases of anti-Larmor and Larmor orbits. However, for the Larmor orbits in the presence of the magnetic field the specific energy ${\cal E}$ for such orbits is larger than for ISCO with $b=0$, while ${\cal E}$ for anti-Larmor orbits decreases with the magnetic field. This means that for a given direction of the magnetic field and direction of the particles motion the behavior of charged particles with opposite charges is qualitatively different. The `anti-Larmor' particles can continue their circular motion and after losing their energy and angular momentum they can reach their ISCO located close to the black hole. Particle with the opposite charge after passing through the radius $6M$ must fall into the black hole. This provides one with quite interesting mechanism of charge separation in the magnetized black holes. In the present paper we discuss only motion of the anti-Larmor particles near magnetized black holes. In order to simplify formulas from now on we omit the indices $\pm$ in the expressions similar to \eq{pm1}--\eq{pm2}.

Figure~\ref{Energy} shows that for anti-Larmor ISCO the specific energy ${\cal E}$ can be arbitrary small. This means that for large value of $b$ the energy $(1-{\cal E})mc^2$, which can be extracted from a charged anti-Larmor particle before it reaches ISCO can be close to the proper energy $mc^2$ of the particle. Thus, the efficiency of the energy extraction in the magnetized black holes can be high and exceed the efficiency of the Kerr black hole.

Anti-Larmor ISCO particles have velocity $v$ (as measured by local rest observer) which faintly depends on the value of the magnetic field (see Figure~\ref{Velocity}). It has the value $1/2$ for both the limits $b\to 0$ and $b\to\infty$. Moving with such velocity the particle returns to the local rest observer after time $2\pi\rho/v$ in his/her local time. Because of the time delay, a distant observer would see this motion as slowed down by his/her clocks. As a result of this effect, the observed angular velocity $\Omega$ at the ISCO tends to zero in the strong field limit $b\to \infty$ (see Figure~\ref{Omega}).

\section{Null rays}

\subsection{Conserved quantities and equations of motion}

A distant observer receives information from an emitter revolving around the black hole by observing its radiation. Two different types of the observations are of the most interest: (1) Study of the images of the emitter orbits and (2) Study of the spectral properties of the observed radiation. The theoretical technics required for these two problems are slightly different. However, in both the cases one needs at first to perform similar calculations. Namely, one needs to integrate equations for the light propagation in the Schwarzschild geometry. This is a well studied problem. Many results concerning ray tracing  as well as the study of the narrow spectral line broadening in the Schwarzschild spacetime can be found in the literature (see, e.g. \cite{FZ} and references therein). Since the magnetic field does not affect the photons propagation one can use similar technics for our problem. However, there are two new features of the problem. Namely, (1) the radius of the emitter can be less than $6M$, the ISCO radius for a neutral article, and (2) even if the charged emitter is at the same orbit as a neutral one, its angular velocity differs from the Keplerian velocity. For this reason one should perform the required calculations and adapt them to a new set-up of the problem.

In this section, we briefly remind main useful formulas concerning the null rays propagation in the Schwarzschild spacetime and fix notations used later in this paper.

Geodesic equation for a null ray is
\be
{D p^{\mu}\over d\lambda}=0\hh g_{\mu\nu}p^{\mu} p^{\nu}=0\, ,
\ee
where $p^{\mu}=dx^{\mu}/d\lambda\equiv \dot{x}^{\mu}$ and $\lambda$ is an affine parameter. For the symmetries \eq{sym1}-\eq{sym3} there exist three commuting integrals of motion
\ba
&&E=-p_{\mu} {\xi}_{(t)}^{\mu}=-p_t=f \dot{t}\, ,\\
&&L_z=p_{\mu} {\xi}_{(\phi)}^{\mu}=p_{\phi}=\rho^2 \sin^2\theta \dot{\phi}\, ,\\
&&L^2=[p_{\mu} {\xi}_{(\phi)}^{\mu}]^2
+[p_{\mu} {\xi}_{x}^{\mu}]^2+[p_{\mu}{\xi}_{y}^{\mu}]^2\nonumber\\
&&=p_{\theta}^2+{p_{\phi}^2\over \sin^2\theta}=
\rho^4(\dot{\theta}^2+\sin^2\theta \dot{\phi}^2)\, .
\ea

Let us remind that, we use the dimensionless quantities. In particular, this means that the `physical' energy is $r_g^{-1} E$.
In what follows, it is convenient to use the following quantities:
\be
\zeta=\rho^{-1}\, , \ \ell_z={L_z\over E}\, ,\  \ell={L\over E}\, , \ \sigma={E\lambda}\, .
\ee

\subsection{Motion in the equatorial plane}

The motion of a ray (as well as the motion of any particle) in the Schwarzschild geometry is planar. One can always choose this plane to coincide with the equatorial plane. For such a choice $p_{\theta}=0$ and $L=|L_z|$. Thus, one has only one conserved quantity, $\ell_z$. To make notations brief we denote it simply as $\ell$.
The equation of motion in the equatorial plane can be written in the following first order form:
\ba
\zeta'&=&-\epsilon \zeta^2 {\cal P}\hh
{\cal P}=\sqrt{1- \ell^2(1-\zeta)\zeta^2}\, ,\\
t'&=&1/(1-\zeta)\hhh \phi'=\ell \zeta^2\hhh  (\ldots)'=d(\ldots)/d\sigma\, .
\ea
For outgoing rays, when $r$ increases along the trajectory, $\epsilon=+1$, and $\epsilon=-1$ for incoming rays. For fixed value of the impact parameter $\ell$ the radial turning point $\zeta_m$ (if it exists) is determined by the condition
\be\n{zm}
(1-\zeta_m)\zeta_m^2=\ell^{-2}\, .
\ee
The evolution of the angle $\phi$ along the trajectory can be found from the following equation:
\be
{d\phi\over d\zeta}=-\epsilon {\ell \over {\cal P}}\, .
\ee

\begin{figure}
  \includegraphics[width=7cm]{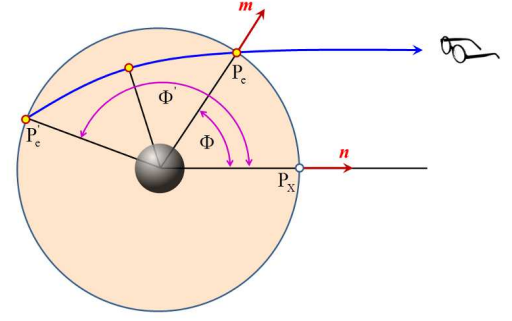}\\
  \caption{Motion of a photon in the equatorial plane. The photon emitted at $P_e$ propagates to a distant observer along a trajectory without radial turning points (a direct ray). The ray emitted at $P'_e$ is indirect. It at first moves to the black hole and only after it passes through a radial turning point it propagates to the distant observer.}\label{Fig1}
\end{figure}

In what follows, we shall use the following function:
\be
B(\ell;\zeta)=\int_0^\zeta {d\zeta \over \sqrt{\ell^{-2}-(1-\zeta)\zeta^2}}
\, .\n{BBB}
\ee
The integral \eq{BBB} can be written in terms of the elliptic function of the first kind $F(x,\alpha)$\footnote{Here we use the definition
\be\nn
F(x,\alpha)=\int^{x}_{0}\frac{d\zeta}{\sqrt{1-\zeta^{2}}\sqrt{1-\alpha^{2}\zeta^{2}}}\,.
\ee
}. One has

\be
B\left(\ell , \zeta \right)=\frac{2\sqrt{2}\ell^{1/3}}{k_{+}}
\left. F\left(X(\zeta),\frac{k_{-}}{k_{+}}\right)\right|_{0}^{\zeta}\, ,
\n{solll3(2)m}
\ee
where,
\ba
&&X(\zeta)=\frac{\sqrt{6}\sqrt{\ell^{2/3}
\left(3\zeta-1\right)
+\sqrt{3}\left(c_{+}+c_{-}\right)}}{3k_{-}}\, ,\\
&&k_{\pm}=\sqrt{\sqrt{3}(c_{+}+c_{-}) \pm i(c_{+}-c_{-})}\, ,\n{factork1}\\
&&c_{\pm}=\left[\frac{\sqrt{3}(\frac{27}{2}-\ell^{2})}{9} \pm \frac{\sqrt{27-4\ell^{2}}}{2}\right]^{1/3}\ . \n{factorc1}
\ea

Consider a ray emitted at the radius $r_e=r_g/\zeta_e$ that reaches the infinity. Denote by $\Phi$ the angle between the direction to the point of emission and the direction to the point of observation. It is easy to show that a null ray can have no more than one radial turning point. The emitted ray either propagates to infinity with monotonic increase of $\rho$, or it at first moves to the smaller value of $\rho$ and only after that goes to infinity. In the former case the bending angle is
\be\n{ba1}
\Phi= B(\ell,\zeta_e)\, .
\ee
In the latter case one has
\be
\Phi= 2 B(\ell,\zeta_m)-B(\ell,\zeta_e)\,.\n{ba2}
\ee
Here $\rho_m=\zeta_m^{-1}$ is a radius of the radial turning point.

\subsection{Integrals of motion and impact parameters}

We are interested in study of propagation of photons emitted by an object revolving around a magnetized black hole in the plane orthogonal to the magnetic field. Using the freedom in the rigid rotations,
it is convenient to choose the spherical coordinates so that this plane coincides with the equatorial plane $\theta=\pi/2$. For such a choice, in the general case, the plane determined by the trajectory of the emitted photon will be tilted with respect to the equatorial one. There still remains freedom in rotation in the $\phi-$direction, preserving the plane connected with the charged particle motion, which we shall fix later.

To derive properties of such photons we write the corresponding equations
\ba
\zeta'&=&-\epsilon \zeta^2 {\cal P}\, ,\n{zeq}\\
\theta'&=&\epsilon_{\theta}\zeta^2 \sqrt{\ell^2-{\ell_z^2\over \sin^2\theta}}\,, \n{teq}\\
\phi'&=&{\ell_z \zeta^2\over \sin^2\theta}\hh {t}'={1\over 1-\zeta} \, .\n{pheq}
\ea

The turning points of the $\theta-$motion, where $\theta'=0$, are determined by the condition $\sin\theta=|\ell_z|/\ell$. Denote these angles by $0< \theta_*\le \pi/2$ and $\pi- \theta_*$. Denote by $\iota$  angle between the normal to the tilted plane and the normal to the equatorial plane. One has
 $\iota=\pi/2-\theta_*$, so that $0\le \iota< \pi/2$ and
\be\n{iota}
\cos\iota ={|\ell_z|\over \ell}\, .
\ee

In what follows, we shall consider rays emitted by a revolving body which propagate to infinity, where an observer is located. In order to characterize asymptotic properties of these rays, which are directly connected with observations one can proceeds as follows. Denote by $\theta_o$ and $\phi_{o}$ the asymptotic angles for the ray trajectory.
The angles of displacement of the photon in the $\theta$ and $\phi$ directions are
$\rho\,d\theta/dt$ and $\rho \sin\theta d\phi/dt$. These angles decrease as $\rho^{-1}$. Multiplying them by $\rho$ and taking the limit $\rho\to \infty$ one obtains the dimensionless impact parameters
\ba
\xi^{\theta}&=&\lim_{\rho\to\infty} \left[\rho^2 {p^{\theta}\over p^{t}}\right]=
\lim_{\zeta\to 0}  \left[\zeta^{-2} {{\theta}'\over t'}\right]\nonumber\\
&=&
\epsilon_{\theta}\sqrt{\ell^2 -{\ell_z^2\over \sin^2\theta_o}}\, ,\n{xit}\\
\xi^{\phi}&=&-\lim_{\rho\to\infty}  \left[\rho^2 \sin\theta {p^{\phi}\over p^{t}}\right]=
-\lim_{\zeta\to 0}  \left[ \zeta^{-2} \sin\theta {{\phi}'\over t'}\right]\nonumber\\
&=& -{\ell_z\over \sin\theta_o}\, .\n{xip}
\ea

Consider a unit sphere with the coordinates $(\theta,\phi)$ and denote a plane tangent to it at the point $(\theta_o,\phi_{o})$ by $\Pi$. We call it the {\em impact plane}. Denote by $\BM{e}_{\theta}$ and $\BM{e}_{\phi}$ unit vectors in $\Pi$ directed along the coordinate lines of $\theta$ and $\phi$, correspondingly. We call the vector
\be\n{iv}
\BM{\xi}=\xi^{\theta}\BM{e}_{\theta}+\xi^{\phi}\BM{e}_{\phi}\,
\ee
the {\em impact vector}. Its norm is
\be
|\BM{\xi}|=\sqrt{(\xi^{\theta})^2+(\xi^{\phi})^2}=\ell\, .
\ee
One also has
\be
\tan\eta\equiv\xi^{\phi}/\xi^{\theta}={\epsilon_{\theta}\cos\iota \ \mbox{sign}(\ell_z)\over \sqrt{\sin^2\theta_o-\cos^2\iota}}\, .
\ee

\subsection{Asymptotic data for null rays}

For study outgoing null rays it is convenient to rewrite the Schwarzschild metric \eq{1d} in the retarded time coordinates
\ba\n{mf1}
ds^2&=&\zeta^{-2} d\tilde{s}^2\, ,\n{cm1}\\
d\tilde{s}^2&=&-\zeta^{2} f du^2 +2du d\zeta +d\Omega^2\, ,\n{cm2}
\ea
where $du=dt-d\rho/f$.
The conformal metric, \eq{cm2}, is especially convenient for describing the asymptotic properties of null rays ar $r\to\infty$. This metric is regular  at the infinity $\zeta=0$, so that the 3D surface $\zeta=0$ with the coordinates $(u,\theta,\phi)$ is nothing but the future null infinity $\mathcal{J}^+$ for our spacetime. Rays with the same asymptotic parameters $(u_{o},\theta_o,\phi_o)$ are asymptotically parallel in the `physical' spacetime with the metric \eq{cm1}. To fix a ray in such a beam one needs two additional parameters, namely the impact vector \eq{iv}. Thus, a point $(u_{o},\theta_o,\phi_o)$ at $\mathcal{J}^+$ together with the impact vector $\BM{\xi}$ uniquely specify a null ray which reaches infinity. We call these five parameters the {\em asymptotic data}.

Equations (\ref{teq}) are equivalent to the following set of equations:
\ba
{d\theta\over d\zeta}&=&-{\epsilon \epsilon_{\theta}\over {\cal P}} {\sqrt{\ell^2-{\ell_z^2\over \sin^2\theta}}}\, ,\n{tteq}\\
{d\phi\over d\zeta}&=&-{\epsilon \ell_z\over {\cal P}\sin^2\theta}\, ,\n{peq}\\
{du\over d\zeta}&=&-{\epsilon \ell^2\over  {\cal P} (1+\epsilon {\cal P})}\, .\n{ueq}
\ea
We remind that for the outgoing ray $\epsilon=+1$.
For given position at ${\cal J}^+$ and the impact vector $\BM{\xi}$ one can determine the integrals of motion $\ell_z$ and $\ell$. For given asymptotic data one can integrate equations (\ref{tteq})--(\ref{ueq}) back in time, from the starting point $\zeta=0$, and to restore the complete null ray trajectory.

\section{Orbit's images}

\subsection{Angular relations}

Denote by $P_e=(t_e,\rho_e,\theta=\pi/2,\phi=\varphi(t_e))$, $\varphi\in(-\pi,\pi]$, an event of the radiation of a quantum by the emitter revolving around the black hole. This quantum is registered by a distant observer $P_o=(u_{o},\zeta_o=0,\theta_o,\phi_o)$. Here, $u_o$ is the moment of the retarded time when the ray arrives to the observer, and $\theta_o$ is the angle between his/her position and the direction orthogonal to the plane $\theta=\pi/2$. We use the freedom of rigid rotations around the axis $Z$ and put the angle $\phi_o$ at the point of observations equal to zero.

\begin{figure}
  \includegraphics[width=8cm]{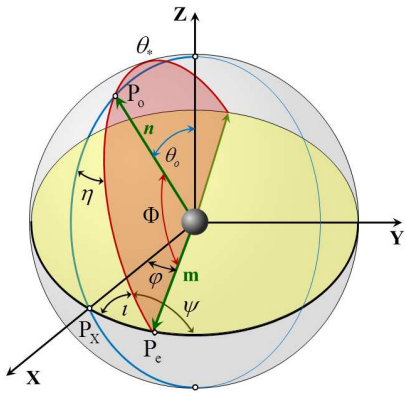}\\
  \caption{Angular definitions.}\label{Fig3}
\end{figure}

For a discussion of the photons trajectories it is convenient to use
a unit round sphere shown in Figure~\ref{Fig3}, which allows one to represent motion of photons and the emitter in the 2D $(\theta,\phi)-$sector. We embed this sphere in a flat 3D Euclidean space, so that a point on the surface of the sphere is uniquely determined by a unit vector with  the origin at the center of the sphere. We call this 2D space an {\em angular space}. Motion of the emitter is represented by the equator of the sphere, while orbits of photons, since they are planar, are represented by large circles.
We use the same letters $P_e$ and $P_o$ as earlier to denote a position of the points of the emission and the observer location in the angular space.

A trajectory of the photon emitted at $P_e$ and arriving to the observer $P_o$ is represented by a large circle, passing through these two points  (see Figure~\ref{Fig3}). We denote by $\Phi$ the angle between the vectors $\vec{n}$ and $\vec{m}$ from the center of the unit sphere to the points $P_o$ and $P_e$, respectively. We call $\Phi$ the {\em bending angle}. It changes from $\pi/2-\theta_o$, when $\varphi=0$ to $\pi/2+\theta_o$ when $\varphi=\pi$.
We call such rays with $\Phi\in[\pi/2-\theta_o,\pi/2+\theta_o]$ {\em primary} to distinguish them from {\em secondary} rays, that make one or more turns around the black hole before they reach the observer. The characteristic property of these rays is that after the emission they move at first below the equatorial plane.
The brightness of the secondary images generated by such rays is greatly suppressed. That is why we do not consider them in our paper.

As one can see from the Figure~\ref{Fig3}, the angle $\theta$ for the primary rays emitted in the interval $\varphi\in [-\pi/2,\pi/2]$ monotonically decreases from the point of emission to $P_o$. For the rays emitted from the other part of the circle the angle $\theta$ at first decreases. It increases after passing through its minimal value (an angular turning point $\theta_*$). Equation (\ref{xit}) implies that in the former case the coordinate of the image on the impact plane $\xi^{\theta}\le0$, while for the latter one $\xi^{\theta}>0$. This means that the image of the part of the emitter trajectory lying in the half-plane with positive $X$ is located in the lower half of the impact plane $\xi^{\theta}<0$, while the part with $X<0$ has the image in the upper half of the impact plane $\xi^{\theta}>0$. Two points with $\xi^{\theta}=0$ are images of the radiation send by the emitter when its crosses the $Y$-axis, where $\varphi=\pm \pi/2$. We denote by $\psi$ an angle between the direction of the photon motion and the velocity of the emitter. The angle $\psi$  is connected with the angle $\iota$ [see \eq{iota}] as follows: $\iota=\pi-\psi$.

A simplest way to find relations between angles which will be used later is to
consider a spherical triangle on a unit sphere. Denote by $A$, $B$ and $C$ its angles, and by $a$, $b$ and $c$ the length of the sides of the triangle, opposite to $A$, $B$ and $C$, respectively. Then, one has
\be
{\sin A\over \sin a}={\sin B\over \sin b}={\sin C\over \sin c}\, ,\n{ang1}
\ee
\be
\cos a=\cos b\cos c+\sin b\sin c \cos A\, .\n{ang2}
\ee
For example, consider the spherical triangle $P_X P_e P_o$ (see Figure~\ref{Fig3}). It has the angles $\pi/2$ (at $P_X$), $\pi-\psi$ (at $P_e$) and $\eta$ (at $P_o$). The length of the sides opposite to the apexes $P_X$, $P_0$, and $P_e$ of this triangle are $\Phi$, $\varphi$, and $\pi/2-\theta_o$, respectively. Using \eq{ang1} and \eq{ang2} one obtains
\be
\cos\Phi=\cos\varphi\sin\theta_o\, ,\
\sin\psi={\cos\theta_o\over \sin\Phi}\, ,\
\sin\eta={\sin\varphi\over \sin\Phi}\, .
\ee

These equations together with the expressions \eq{xit}--\eq{iv} allow one to determine the impact vector $\vec{\xi}$ in terms of the total angular momentum $\ell$, the inclination angle $\theta_o$ and the position angle $\varphi$ of the emitter
\ba
\xi^{\phi}&=&\frac{\ell\sin\varphi}{\sin\Phi}\, ,\n{map1}\\
\xi^{\theta}&=&\frac{\ell\cos\varphi\cos\theta_{o}}{\sin\Phi}\, . \n{map2}
\ea
One also has
\ba
\ell_z&=&-{\ell \sin\theta_o \sin\varphi
\over \sin\Phi}\, ,\n{map3}\\
\sin\Phi&=&\sqrt{\sin^2\varphi +\cos^2\theta_o\cos^2\varphi}\, .
\ea
These relations, besides the inclination angle $\theta_o$ of the orbit  and angular position of the emitter, $\varphi$, contain only one unspecified parameter $\ell$.

\subsection{Map between equatorial and  impact planes}

\begin{figure}
  \includegraphics[width=8cm]{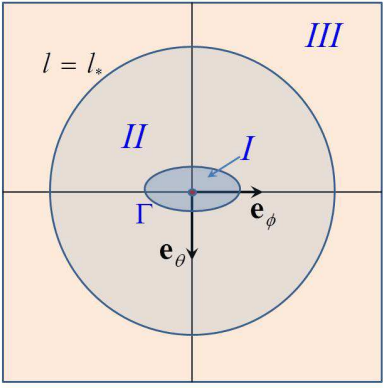}\\
  \caption{Impact plane}\label{Fig4}
\end{figure}

Let us now consider the radial equation (\ref{zeq}). For $\ell > \ell_*= 3\sqrt{3}/2$ (domain $III$ in Figure~\ref{Fig4}) the ray trajectory has a radial turning point. Such a ray reaches the minimal radius that is always greater than $3r_g/2$, and after goes to infinity again. These rays always cross the equatorial plane in the black hole exterior.
For $\ell < \ell_*= 3\sqrt{3}/2$ the ray trajectory does not have a radial turning point, so that being traced back from ${\cal J}^+$ they reach the horizon of the black hole.  The radial ray, with $\ell=0$, does not cross the equator before it  enters the black hole. The same property have rays reaching ${\cal J}^+$ in some vicinity of this ray with  sufficiently small impact parameter $\ell$ (domain $I$ in Figure~\ref{Fig4}).
They enter the black hole {\em before} they cross the equatorial plane.
The rays with impact parameters outside the region $I$
cross the equatorial plane first. In both the cases, $I$ and $II$ one must put $\epsilon=+1$, since the radial turning points are absent.

\begin{figure}
  \includegraphics[width=5cm]{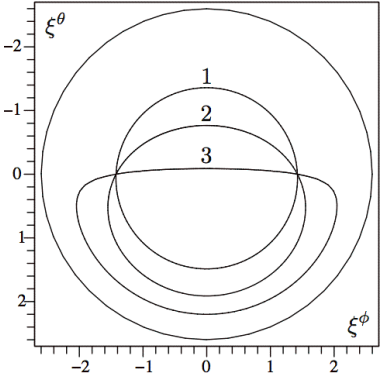}\\
  \caption{The boundary $\Gamma$ between the regions $I$ and $II$ for different values of the observer's angle $\theta_{o}$: Curve 1: $\theta_{o}=5^{o}$; Curve 2: $\theta_{o}=45^{o}$; Curve 3: $\theta_{o}=85^{o}$. A large circle is a curve $|\BM{\xi}|=\ell_*=3\sqrt{3}/2$.}\label{Fig3a}
\end{figure}

Using \eq{ba1} one can find the boundary between the regions $I$ and $II$ from the following relation:
\be
B(\ell,1)=\Phi(\varphi,\theta_o)=\arccos(\cos\varphi\sin\theta_{o})\, .
\ee
(Let us remind that the rays in this domain do not have a radial turning point.) Figure~\ref{Fig3a} presents solutions of this equation for different values of the inclination angle $\theta_o$.

Let us summarize this part. For a fixed position of the observer at infinity $(\theta_o,\phi_o=0)$ and the moment of arrival of the rays $u_o$ there exist a one-to-one correspondence between the region $II\bigcup III$ of the impact plane and the region of the equatorial plane, located outside the black hole. We call this map $\Psi$
\be
\Psi: \BM{\xi} \to (\zeta<1,\theta=\pi/2,\phi)\, .
\ee

\subsection{Direct and indirect rays}

\begin{figure}
  \includegraphics[width=6cm]{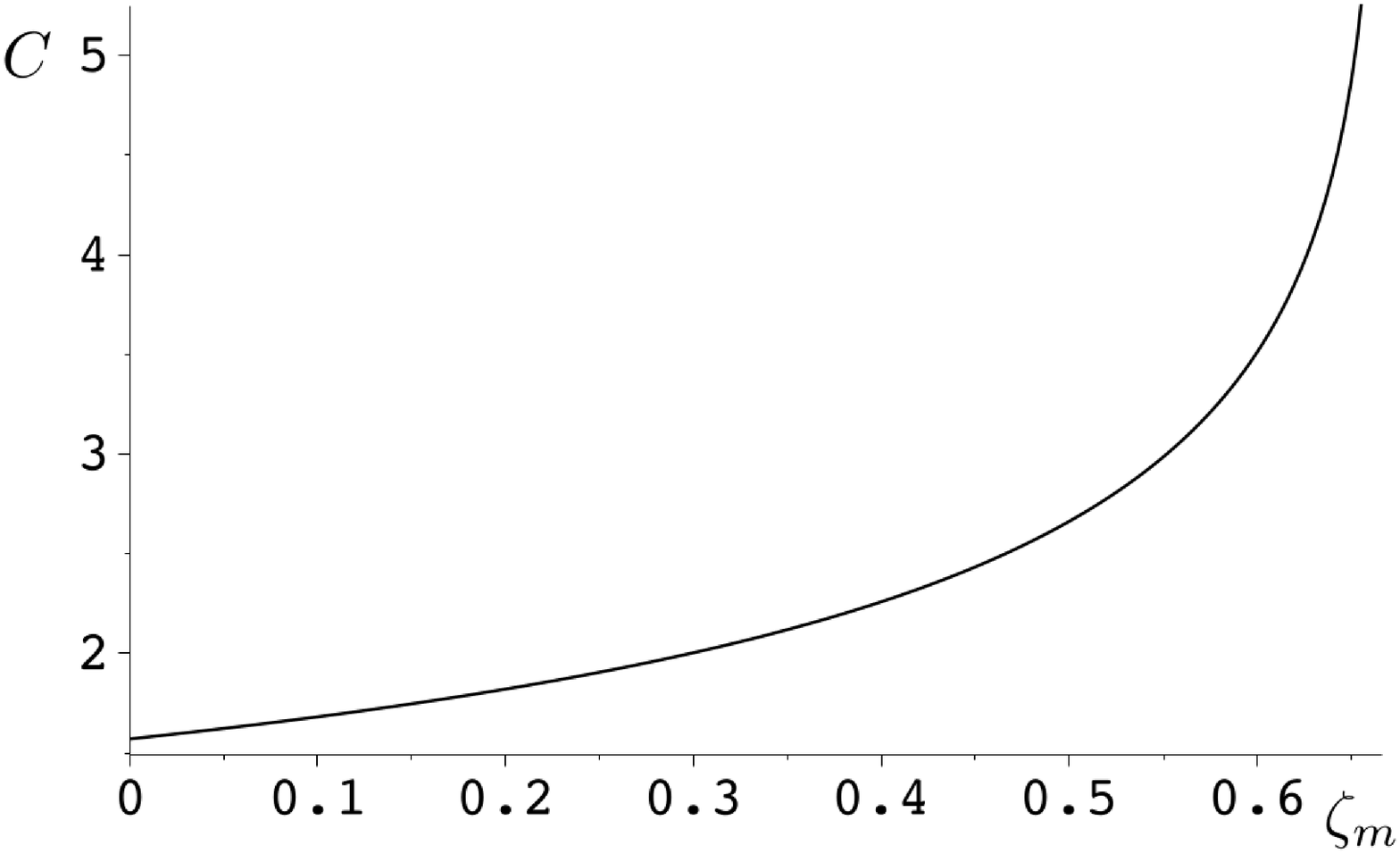}\\
  \caption{Function $C(z)$. It monotonically grows with $z$ from $\pi/2$ at $z=0$  and becomes infinite at $z=2/3$.}\label{BBZ}
\end{figure}

Consider a ray connecting a point $P_e$ on the equatorial plane and a distant observer $P_o$. If such a ray does not have a radial turning point we call it {\em direct}. In the opposite case we call it an {\em indirect} ray (see Figure~\ref{Fig1}). If the radius of the emitter orbit is small enough all the rays from it to the distant observer are direct. For a larger radius of the orbit there exist such a value $|\varphi_*|\ge \pi/2$ of the angle $\varphi$ that for the part $-\varphi_*<\varphi<\varphi_*$ one has only direct rays, while for  $\pi >\varphi>\varphi_*$ and $-\pi<\varphi<-\varphi_*$ the rays are indirect. We denote the critical inverse radius which separates these two cases by $\zeta_*$. For $\zeta_e=\zeta_*$ the radial turning point is located at $\varphi_*=\pi$ on the equatorial plane. For this case one has
\be
\Phi=\Phi_*(\theta_o)=\pi/2+\theta_o\, .
\ee
The critical value $\zeta_*$ is a function of the inclination angle $\theta_o$. To find it let us denote by $C(z)$ the following integral
\be
C(z)=\int_0^{z} {d\zeta \over \sqrt{(1-z)z^2-(1-\zeta)\zeta^2}}\, .
\ee
By change of the variable $\zeta=z(1-y^2)$ this integral can be rewritten in the form
\be\n{CC}
C(z)=2\int_0^{1}{dy\over \sqrt{Z}}\, ,\ 
Z=2-y^2-3z+3z y^2-z y^4\, .
\ee
Figure~\ref{BBZ} show a plot of this function.

\begin{figure}
  \includegraphics[width=7cm]{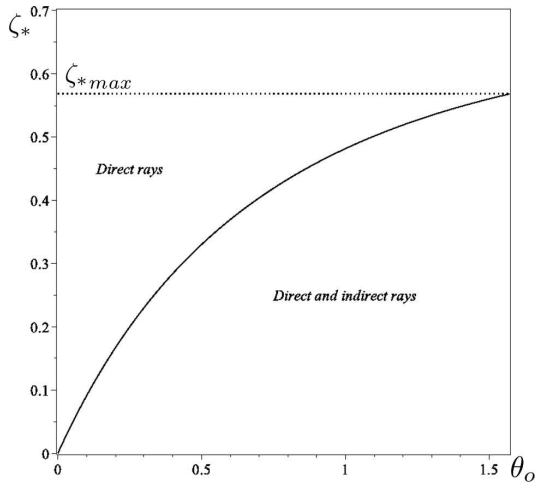}\\
  \caption{The critical inverse radius $\zeta_*$ as a function the inclination angle $\theta_o$.}\label{CRIT}
\end{figure}

The function $\zeta_*(\theta_o)$ is determined by the relation
\be
C(\zeta_*)=\Phi_*=\pi/2+\theta_o\, .
\ee
The plot of $\zeta_*(\theta_o)$ is shown in Figure~\ref{CRIT}. $\zeta_*$ monotonically increases from 0 (at $\theta_o=0$) to its maximal value at $\theta_o=\pi/2$ equal to
\be
\zeta_{*,max}\approx0.5680820870\, .
\ee

Consider now a circular orbit with $\zeta_e<\zeta_*$. The following equation determines an angle $\varphi_*$ on such an orbit which separates its two parts (with direct and indirect rays):
\be
\varphi_*=\arccos\left( {\cos(C(\zeta_e))\over \sin\theta_o}\right)\, .
\ee

It is convenient to combine the relations \eq{ba1} and \eq{ba2}. We introduce the functions $B_{\pm}(\ell,\zeta_e)$ which is defined as follows:
\be
B_{+}(\ell,\zeta_e)=B(\ell,\zeta_e)\, ,\ \
B_{-}(\ell,\zeta_e)=2C(\zeta_m(\ell))-B(\ell,\zeta_e)\, ,
\ee
where $\zeta_m$ is defined by \eq{zm}, $(1-\zeta_m)\zeta_m^2=\ell^{-2}$, and function $C$ is defined by \eq{CC}. Note that for numerical computations of $B_{-}(\ell,\zeta_e)$ it is more convenient to consider $\zeta_{m}$ as a parameter.

The following equation:
\be\n{BBF}
B_{\pm}(\ell,\zeta_e)=\Phi\equiv\arccos(\cos\varphi\sin\theta_0)\, ,
\ee
establishes a relation between position (angle $\varphi$) of the emitter on the orbit with the inverse radius $\zeta_e$ and the angular momentum $\ell$ of the photon that reaches a distant observer with the inclination angle $\theta_o$. In this relation one need to choose sign $+$ for a direct trajectory and $-$ for an indirect one.

The corresponding image on the impact plane can be found by using \eq{map1} and \eq{map2}. By integrating equation (\ref{ueq}) one obtains a relation between the time of emission, $t_{e}$, and the retarded time of observation, $u_{o}$, at ${\cal J}^+$.

\begin{figure}
\includegraphics[width=6cm]{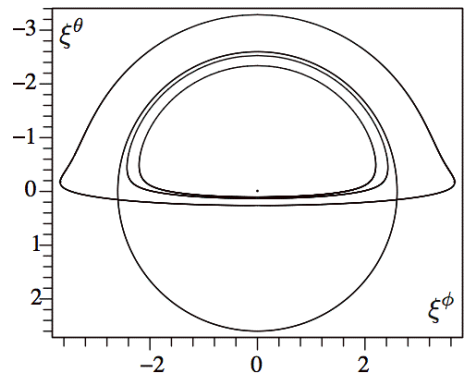}\\
\caption{Images of some orbits corresponding to $\theta_{o}=85^{o}$. The innermost curve is the image of the $\zeta_{e}=5/6$ orbit, it is formed by direct null rays. The next curve is the image of the $\zeta_{e}=2/3$ orbit, it is formed by direct null rays as well. And the outermost curve is the image of the $\zeta_{e}=1/3$ orbit. The lower part of the curve is formed by direct null rays. The upper part of the curve is formed by indirect rays. The circle represents the rim of the black hole shadow.}\label{F11}
\end{figure}

In conclusion of this section let us give examples of the images of orbits on the impact plane. These images for the inclination angle $\theta_o=85^{o}$ and the inverse radius of the orbit equal to $\zeta_{e}=5/6, 2/3,1/3$
are shown in Figure~\ref{F11}.

\section{Spectral broadening}

\subsection{Photon momentum and conserved quantities}

In what follows, we use the following orthonormal tetrad at the point of emission $P_e$:
\be\n{TTT}
\{ \BM{e}_t,\BM{e}_{\rho}, \BM{e}_{\Phi}, \hat{\BM{e}}\}\, .
\ee
The first of the vectors, $\BM{e}_t$, is in the direction of the Killing vector $\BM{\xi}_{(t)}$. The second vector $\BM{e}_{\rho}$ is along the radial direction, while the last two vectors are tangent to the surface $t=$\,const and $\rho=$\,const. We choose $\BM{e}_{\Phi}$ to lie in the photon orbit plane and directed from the point of emission $P_e$ to the point of observation $P_o$. The last vector $\hat{\BM{e}}$ is uniquely defined by the condition that the tetrad \eq{TTT} is right-hand oriented. The unit vector $\BM{e}_{\phi}$ in the equatorial plane and tangent to the orbit of the emitter can be written in the form
\be\n{ephi}
\BM{e}_{\phi}=-{1\over \sin{\Phi}} (\sin\theta_o\sin\varphi \, \BM{e}_{\Phi}+\cos\theta_o \, \hat{\BM{e}})\,,
\ee
where
\be
\hat{\BM{e}}=\frac{1}{\sin\Phi}(\sin\varphi\cos\theta_{o},-\cos\varphi\cos\theta_{o},-\sin\varphi\sin\theta_{o})\,.
\ee

Consider a photon with the impact parameter $\ell$. Denote by $\Gamma$ the plane of its orbit, and by $\BM{\xi}_{(\Phi)}$ the Killing vector generating rotations preserving $\Gamma$. Then the momentum of the photon at the moment when it pass the radius $\rho$ can be written in the form
\be
\BM{p}=\nu (\BM{\xi}_{(t)}+a\BM{e}_{\rho}+b\BM{\xi}_{(\Phi)})\, .
\ee

One has
\ba
\omega_{o}&=&-(\BM{p},\BM{\xi}_{(t)})=\nu f\, ,\\
L&=&(\BM{p},\BM{\xi}_{(\Phi)})=\nu b \zeta^{-2}\, .
\ea
Thus, one has
\be
\ell={L\over \omega_{o}}={b\over \zeta^2 f}\, .
\ee
The quantity $\omega_{o}$ is the frequency of the photon at infinity, measured in $r_g^{-1}$ units.
Using these relations and the property $\BM{p}^2=0$ one finds
\be\n{ppp}
\BM{p}=\omega_{o} ({1\over \sqrt{f}}\BM{e}_{t}+
{1\over \sqrt{f}}{\cal P}\BM{e}_{\rho}+\ell \zeta \BM{e}_{\Phi})\,,
\ee
where, as earlier, ${\cal P}=\sqrt{1-\ell^2 \zeta^2 f}$.

\subsection{Red-shift factor}

Using \eq{ppp} one finds the emitted frequency $\omega_e$
\be
\omega_e=-p_{\mu}u^{\mu}=\gamma \omega_o \left[{1\over \sqrt{f}} -v\ell \zeta (\BM{e}_{\phi},\BM{e}_{\Phi})\right]\, .
\ee
Using \eq{ephi} one can write
\be\n{xipP}
(\BM{e}_{\phi},\BM{e}_{\Phi})=-{\sin\varphi \sin\theta_o\over \sin\Phi}\, .
\ee
Combining these results we obtain the following relation between the emitted frequency $\omega_e$ and the frequency $\omega_{o}$ registered by a distant observer
\be\n{f7}
\omega_{o}=\omega_{e}\frac{\sin\Phi\,
\sqrt{f_e-\Omega^{2}\rho_e^{2}}}{(\sin\Phi+\ell\Omega\sin\theta_{o}\sin\varphi)}\,.
\ee

We denote the ratio $\omega_e/\omega_o$ by $\alpha$
\be\n{AL}
\alpha=\gamma_{e}\left(1+{\ell\Omega \sin\varphi\sin\theta_o\over \sin\Phi}\right)\, .
\ee

\subsection{Solid angle at the emitter}

We remind that the photon orbit is planar. We call the corresponding plane a {\em photon plane}. We choose a reference photon emitted to the distant observer. In order to find the solid angle of the emitted photons that pass through the `telescope' aperture we fix the position of the emitter and consider a bundle of emitted photons with momenta that slightly differ from the momentum $\BM{p}$ of the reference photon. To determine the bundle we consider two types of variations, which do not involve the trivial conformal variation of $\omega_{o}$. The first one is the variation $\delta\ell$ of the angular momentum $\ell$ which does not change the photon plane
\be\n{dpl}
\delta_{\ell}\BM{p}=\omega_o \delta\ell \BM{Z}\hh
\BM{Z}=-{\ell \zeta^2 \sqrt{f}\over {\cal P}}\BM{e}_{\rho}+\zeta \BM{e}_{\Phi}\, .
\ee
The second variation $\delta\psi$ changes the position of the photon plane and slightly rotates it around the direction to the emitter by an angle $\delta\psi$.
As a result of this rotation the vectors $\BM{e}_{\Phi}$ and $\hat{\BM{e}}$ are transformed as follows
\be\n{dpp}
\delta_{\psi}\BM{e}_{\Phi}=\hat{\BM{e}}\delta \psi\hh
\delta_{\psi}\hat{\BM{e}}=-\BM{e}_{\Phi}\delta \psi\, .
\ee
Hence
\be\n{ppsi}
\delta_{\psi}\BM{p}=\omega_o \ell\zeta_e \hat{\BM{e}}\delta \psi\, .
\ee

To find the solid angle $\Delta\Omega_e$ we shall use the relation (\ref{EEEE})
and write it in the form
\be
\BM{\cal A}=\pm \Delta\Omega_e \BM{E}\, ,
\ee
where $\BM{E}$ is a unit rank-4 totally skew-symmetric tensor and
\be\n{dpA}
\BM{\cal A}=\omega_e^{-3} \BM{u}\wedge\BM{p}\wedge\delta_{\ell}\BM{p}\wedge\delta_{\psi}\BM{p}\, .
\ee
Using equations (\ref{dpl}), (\ref{dpp}) and (\ref{dpA}) one obtains
\ba
\BM{\cal A}&=&\gamma \ell\zeta_e\left({\omega_o\over \omega_e}\right)^3\delta\ell\ \delta\psi\  ( \BM{\cal B}+v \BM{\cal C})\, ,\\
\BM{\cal B}&=&\omega_o^{-1}\BM{e}_t\wedge\BM{p}\wedge \BM{Z}\wedge \hat{\BM{e}}\, ,\\
\BM{\cal C}&=&\omega_o^{-1}\BM{e}_{\varphi} \wedge\BM{p} \wedge\BM{Z} \wedge \hat{\BM{e}}\, .
\ea
Simple calculations give
\be
\BM{\cal B}={\zeta_e\over \sqrt{f_e}{\cal P}}\BM{E}\hh
\BM{\cal C}={\ell \zeta_e^2 \sin\theta_o\sin\varphi\over {\cal P}\sin\Phi}\BM{E}\, ,
\ee
where
\be
\BM{E}=\BM{e}_t\wedge\BM{e}_{\rho}\wedge\BM{e}_{\Phi}\wedge \hat{\BM{e}}\, .
\ee
Thus one has
\be\n{SSS}
\Delta\Omega_e={\ell \zeta_e^{2} \over \alpha^2 {\cal P}}\delta\ell\  \delta\psi\, .
\ee

\subsection{Spectral broadening}

If there is no caustics, all the photons of the bundle emitted in the solid angle $\Delta\Omega_e$ propagate until they meet the observer's device, which we call a `telescope'. We assume that it is located at the radius $\rho_o$, its aperture is $A$ and it is oriented orthogonal to the bundle of photons. The variation $\delta\ell$ changes the angle $\Phi$ at the point of observation by the value
\be
\delta_{\ell}\Phi=\Phi' \delta\ell\hh
\Phi'\equiv {dB_{\pm}(\ell;\zeta_e)\over d\ell}\, .
\ee
The other variation  is the rotation of the photon's plane around  the direction to the emitter by the angle $\delta\psi$. Under this transformation a point with a fixed value $\Phi$ on the photon's plane is shifted by the angle
\be
\delta\chi=\sin\Phi\,\delta\psi\, .
\ee
in the direction orthogonal to it.
Thus, the area of the bundle of the photons emitted in the solid angle \eq{SSS} on the `screen' orthogonal to the bundle and located at the radius $\rho_o$ is
\be
A= \rho_o^2 \delta_{\ell}\Phi\,\delta\chi=\rho^2_o \Phi'\sin\Phi\,\delta\ell\, \delta\psi\, .
\ee
If instead of the `screen' one uses a `telescope' one can identify $A$ with its aperture.

We denote by ${\cal N}\Delta\tau_e$ the total number of the photons emitted during the proper time $\Delta\tau_e$.
A part of these photons $\Delta_e/(4\pi)$. which is emitted in the solid angle $\Delta_e$  
reaches the aperture of the `telescope' during the corresponding time interval $\Delta t_o$ at the point of the observation. Thus, one has
\ba\n{nooo}
&&{dN_o\over dt_o}={d\tau_e\over d t_o} {\cal N}{\Delta_e\over 4\pi} ={\cal C}{d\tau_e\over dt_o} {\ell \zeta_e^{2}\over \alpha^2 {\cal P}\Phi'\sin\Phi}\, ,\\
&&{\cal C}={{\cal N} A\over 4\pi \rho_o^2}\, .
\ea
$dN_o/ dt_o$ is the number of registered photons per a unit time at the point of observation.
The quantity ${\cal C}$, which enters \eq{nooo} has a simple meaning. Consider a flat spacetime and an emitter at rest. Then ${\cal C}$ is the number of particles registered per a unit time by the observer located at the distance $\rho_o$ from the emitter, provided the aperture of his/her `telescope' is ${A}$. This quantity for a fixed distance $\rho_o$ is just a common factor in \eq{nooo} and similar expressions and it does not depend on details of the emitter's motion. For this reason it is convenient to define new quantities, such as
\be
n_{t_o}={\cal C}^{-1}{dN_o\over dt_o}\hh n_{t_e}={\cal C}^{-1}{dN_e\over dt_e}\, .
\ee
In such a case we say that we are using the {\em Newtonian normalization}.

The above equations allow one to find how the number of the observed quanta depends on the time $t_o$. Instead of this one may ask how observed quanta are distributed over the observed frequency $\omega_o$. In the latter case it is convenient to introduce the spectral distribution of the observed quanta\footnote{One can arrive to the same spectral function by assuming that instead of a single ion, there exist many of such ions at the circular orbit of the same radius $\rho_e$. In such a case, an averaging over the angle $\varphi_e$ is effectively equivalent to the integrating (averaging) over the arrival time $u_o$. So that, one again arrives to the same spectral function $n_{\omega_o}$.}.

\be\n{omom}
n_{\omega_o}= {\Omega\over 2\pi}{n_{t_o}\over |d\omega_o/dt_o|}\, .
\ee
Here we introduce an additional factor $\Omega/2\pi$ which requires an explanation. The observed frequency $\omega_o$ is a periodic function of $t_o$ with the period $T_o=2\pi/\Omega$. This is a time of the complete revolution of the emitter as measured at infinity. As we shall see later, the frequency $\omega_o$ changes in some interval $[\omega_{min},\omega_{max}]$ and  in this interval there exist two branches of the function $\omega_{o}(t_o)$: in the first branch $d\omega_o/dt_o>0$, while in the second one $d\omega_o/dt_o<0$. Denote by $N_o$ the following quantity
\be\n{NPh}
{N}_o=\oint_{\omega_o}n_{\omega_o}|d\omega_o|\, ,
\ee
where the integral is taken over both the branches. This gives the total number of photons received  by the observer during one period of revolution of the emitter divided by the period $T_o$. In order to provide this useful normalization we included the factor $\Omega/2\pi$ in \eq{omom}.

Using \eq{nooo} one obtains
\be\n{SPEC}
n_w\equiv \omega_e n_{\omega_o}= { \Omega\over 2\pi}\left| {d\alpha\over d\tau_e}\right|^{-1} {\ell \zeta_e^2\over \Phi' {\cal P}\sin\Phi}\, .
\ee
Let us now obtain an expression for the time derivative of $\alpha$ which enters \eq{SPEC}. First let us notice that $d / d\tau_e=\gamma(d /dt_e)$, so that one has
\be
{d\alpha\over d\tau_e}={\gamma}^2 \Omega \sin\theta_o \dot{Q}\, ,
\ee
where a dot denotes a derivative with respect to time $t_e$, and
\be
Q={\ell \sin\varphi\over \sin\Phi}\, .
\ee
Simple calculations give $\dot{\varphi}=\Omega$ and
\ba
\dot{Q}&=&{\dot{\ell} \sin\varphi\over \sin\Phi}+{\ell \cos\varphi\over \sin\Phi}\dot{\varphi}-{\ell \sin\varphi\cos\Phi\over \sin^2\Phi}\dot{\Phi}\, ,\\
\dot{\ell}&=&{\sin\varphi \sin\theta_o\over \Phi'\sin\Phi}\dot{\varphi}\, ,\\
\dot{\Phi}&=&{\sin\varphi \sin\theta_o\over \sin\Phi}\dot{\varphi}\, .
\ea
Thus one obtains
\be
\dot{Q}=\Omega \left( {\ell \cos\varphi \cos^2\theta_o\over \sin^3\Phi}{\bf +}
{\sin^2\varphi \sin\theta_o\over \Phi' \sin^{2}\Phi}\right)\, .
\ee
and
\be\n{AAA}
{d\alpha\over d\tau_e}={\Omega^2\sin\theta_o \zeta_e^2\over ([1-\zeta_e]\zeta_e^2-\Omega^2)}
\left( {\ell \cos\varphi \cos^2\theta_o\over \sin^3\Phi}+
{\sin^2\varphi \sin\theta_o\over \Phi' \sin^{2}\Phi}\right)\, .
\ee

Let us remind that we use the dimensionless quantities obtained by the rescaling which involves the gravitational radius $r_g$ of the black hole. However, the quantity ${\cal C}\omega_e^{-1}$ is scale invariant. We denote by $w =\omega_o/\omega_e=\alpha^{-1}$. Then $n_w$ given by \eq{SPEC} is the scale invariant quantity. We call it a {\em spectral function}. The total number of quanta $N_o$ (in the Newtonian normalization) defined by \eq{NPh} is
\be\n{NPh1}
{N}_o=\oint_{w}n_{w}|dw|\, .
\ee

\subsection{General properties of spectral functions}

Before presenting the results of numerical calculations, let us discuss some general expected properties of the spectral function $n_w$ given by equations (\ref{SPEC}) and (\ref{AAA}). We rewrite the expression \eq{AL} in the form
\ba
&&\alpha=\gamma [1+\Omega Z(\varphi)]\hh Z(\varphi)=\ell(\varphi)\hat{Z}\, ,\\
&&\hat{Z}= \pm{\sqrt{\sin^2\theta_o-\cos^2 \Phi}\over \sin \Phi}\, .
\ea
$Z(\varphi)$ is a periodic function of the angle $\varphi$ with the period $2\pi$. At the points $\varphi=0$ and $\pm\pi$ one has $|\cos\Phi|=\sin\theta_o$, so that $Z$ vanishes at these points. Moreover, the function $\hat{Z}$ is antisymmetric with respect to the reflection at $\varphi=0$. Since $\ell$ is a symmetric function of $\varphi$ with respect to the reflection $\varphi\to-\varphi$, the function $Z(\varphi)$ has the maximum $Z_m$ at $\varphi_m\in(0,\pi)$, and the minimum $-Z_m$ at $\varphi_m$. Near these points,  assuming that the function $\ell(\varphi)$ is smooth, one has ($\beta>0$)
\be
Z\sim \pm(Z_m -{1\over 2}\beta (\varphi\mp\varphi_m)^2)\, .
\ee
Since $\varphi=\Omega t_e$ one also has
\be\n{ZE}
{dZ\over dt_e}\sim \mp \beta \Omega (\varphi\mp\varphi_m)\, .
\ee

\begin{figure}[htb]
\begin{center}
\includegraphics[width=5cm]{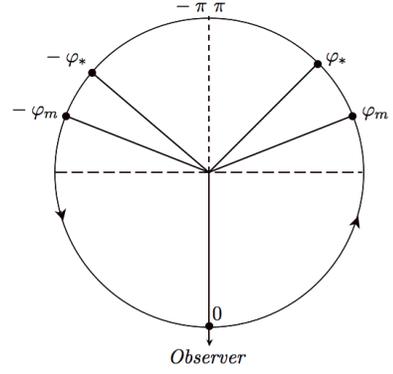}\\
\caption{Diagram illustrating orbit of the emitter. The arrows show the direction of the emitter's motion ($\Omega>0$). For the emitter located in the right semicircle, $\varphi\in[0,\pi]$, photons have Doppler red-shift and for the emitter located in the left semicircle, $\varphi\in(-\pi,0]$, photons have Doppler blue-shift. The spectral function diverges at $\varphi=\varphi_{m}$ and $\varphi=-\varphi_{m}$, where $|\varphi_{m}|>\pi/2$. The portion of the orbit corresponding to indirect null rays is defined by the angle $|\varphi|>\varphi_{*}$}\label{F12}
\end{center}
\end{figure}

Consider now $w=\alpha^{-1}$. This quantity is periodic function of $\varphi$. Denote
\be\n{ggg}
w_{\pm}=\gamma^{-1} {1\over 1\mp\Omega Z_m}\, .
\ee
$w_+$ is the maximal observed frequency of photons. Such photons come  from the emitter when it is at $-\varphi_m$. Similarly, $w_-$ is the minimal observed frequency and the corresponding photons are emitted at $\varphi_m$ (see Figure~\ref{F12}). At these frequencies the spectral function has peaks. The position of the emitter $\varphi=\Omega t_e$ is a regular (linear) function of time everywhere, including the points where the frequency $w$ reaches its extrema and hence $dw/dt_e=0$ at these points. When one transforms the rate of emission to the spectrum, one multiplies the former by the factor $(d\alpha/d\tau_e)^{-1}$. This is the origin of the spectrum peaks. Using \eq{ZE} it is easy to see that $|d\alpha/d\tau_e|\sim |w-w_{\pm}|^{1/2}$ near these points. So that the spectral divergence at the peaks is evidently integrable. It should be so since the total number of photons emitted during one period of the revolution is finite.

It is convenient to introduce a parameter
\be\n{width}
\Delta=  {2(w_+-w_-)\over (w_++w_-)}=2\Omega Z_m
\ee
which determines the width of the spectrum. Figure~\ref{Omega} shows that $\Omega$ at ISCO for the anti-Larmor motion decreases when the magnetic field grows, and $\Omega\to 0$ when $b\to\infty$. Thus one can expect that in the presence of the magnetic field the width parameter decreases. 
One can also conclude from \eq{ggg} that the parameter $\gamma^{-1}$ controls the general redshift of the spectra.
Let us make one more remark concerning the properties of the spectral functions. The radiation from the emitter at $-\varphi_m$ has the maximal Doppler blueshift, while at $\varphi_m$ it has maximal redshift. One can expect that because of the relativistic beaming effect the number of particles with the frequency $w_+$ should be larger than with the frequency $w_-$. This implies that the spectral function should be asymmetric with respect to its minimum, and the peak at $w_+$ must be more profound than the peak at $w_-$.  To characterize the asymmetry of the spectral function with respect to its minima we introduce the {\em asymmetry parameter}:
\be
\delta=\frac{w_{+}+w_{-}-2w_{0}}{(w_{+}-w_{-})}\,,
\ee
where $w_{0}$ corresponds to the minima of a spectral function.

Let us remind also that the obtained spectrum was calculated for a single orbit with a fixed radius. If a radiating domain is a ring of a finite width, one should integrate the spectrum over the radius $\rho_e$ with a weight proportional to the density of the matter of Iron ions in such a ring. After this the infinite peaks disappear and the spectrum would be regularized.

\section{Results}

\begin{figure}[htb]
\begin{center}
\includegraphics[width=5cm]{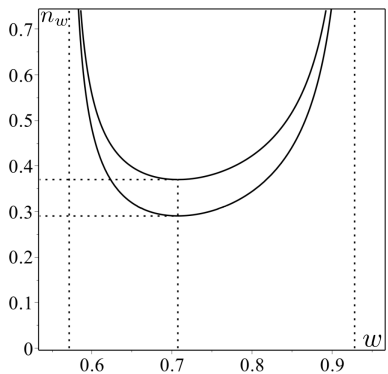}
\caption{Spectral function for ISCO, $b=0$, at $\zeta_{e}=1/3$. The inclination angle is $\theta_o=30^{\circ}$.
The angular velocity of the emitter is $\Omega=0.136$ and its specific energy is ${\cal E}=0.943$. The spectrum has peaks at $w_{-}= 0.571$ ($\varphi_{m}= 97^{o}85'$) and at $w_{+}=0.928$ (at $-\varphi_{m}$).
The minimal values ($0.290$ and $0.370$) of $n_{w}$ for the two spectral branches are at $w_{0}=0.707$. The width parameter is $\Delta=0 .476$ and the asymmetry parameter is $\delta= 0.236$. One also has $N_o=0.285$.
}\label{ISCO1330}
\end{center}
\end{figure}

\begin{figure}[htb]
\begin{center}
\includegraphics[width=5cm]{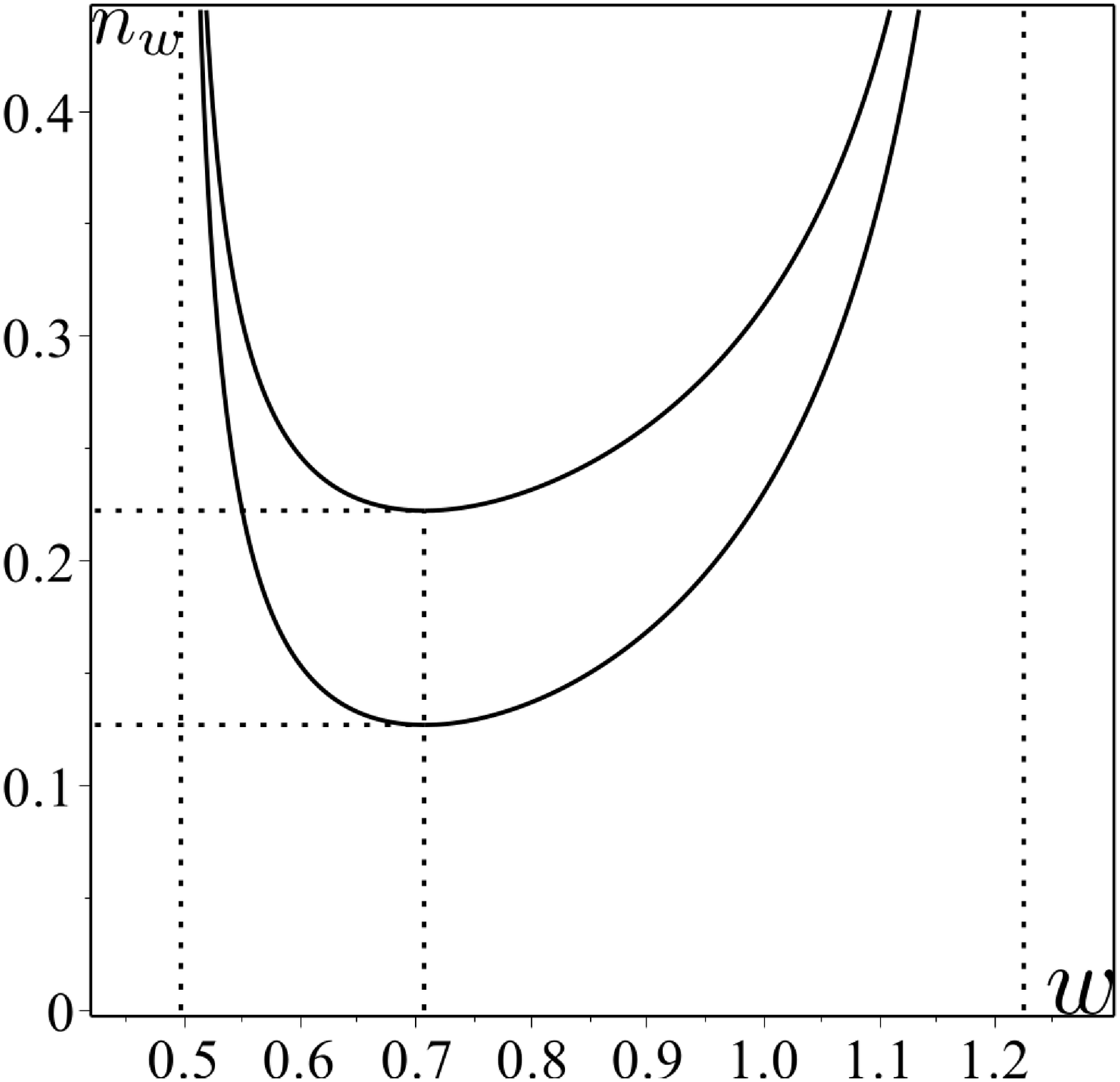}
\caption{Spectral function for ISCO, $b=0$, at $\zeta_{e}=1/3$. The inclination angle is $\theta_o=60^{\circ}$.
The angular velocity of the emitter is $\Omega=0.136$ and its specific energy is ${\cal E}=0.943$. The spectrum has peaks at $w_{-}=0.497$ ($\varphi_{m}= 107^{o}9'$) and at $w_{+}= 1.224$ (at $-\varphi_{m}$).
The minimal values ($0.127$ and $0.222$) of $n_{w}$ for two spectral branches are at $w_{0}=0.707$. The width parameter is $\Delta= 0.845 $ and the asymmetry parameter is $\delta=0.423$. One also has $N_o=0.356$.
}\label{ISCO1360}
\end{center}
\end{figure}

\begin{figure}[htb]
\begin{center}
\includegraphics[width=5cm]{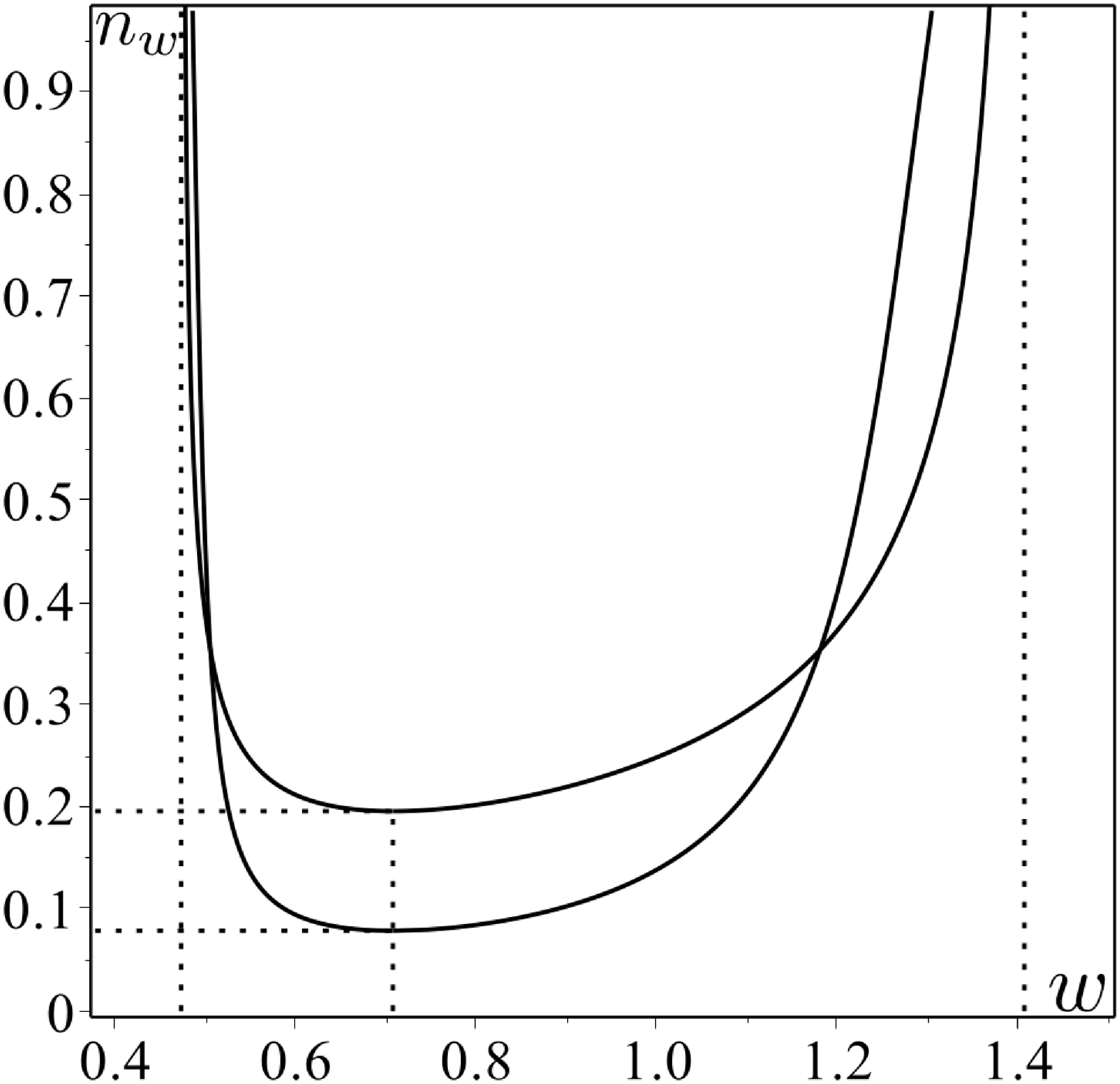}
\caption{Spectral function for ISCO, $b=0$, at $\zeta_{e}=1/3$. The inclination angle is $\theta_o=85^{\circ}$.
The angular velocity of the emitter is $\Omega=0.136$ and its specific energy is ${\cal E}=0.943$. The spectrum has peaks at $w_{-}= 0.472$ ($\varphi_{m}= 118^{o}4'$) and at $w_{+}= 1.407$ (at $-\varphi_{m}$).
The minimal values ($0.079$ and $0.196$) of $n_{w}$ for two spectral branches are at $w_{0}=0.707$. The width parameter is $\Delta=0.995$ and the asymmetry parameter is $\delta= 0.498$. One also has $N_o=0.397$.}\label{ISCO1385}
\end{center}
\end{figure}

\begin{figure}[htb]
\begin{center}
\includegraphics[width=5cm]{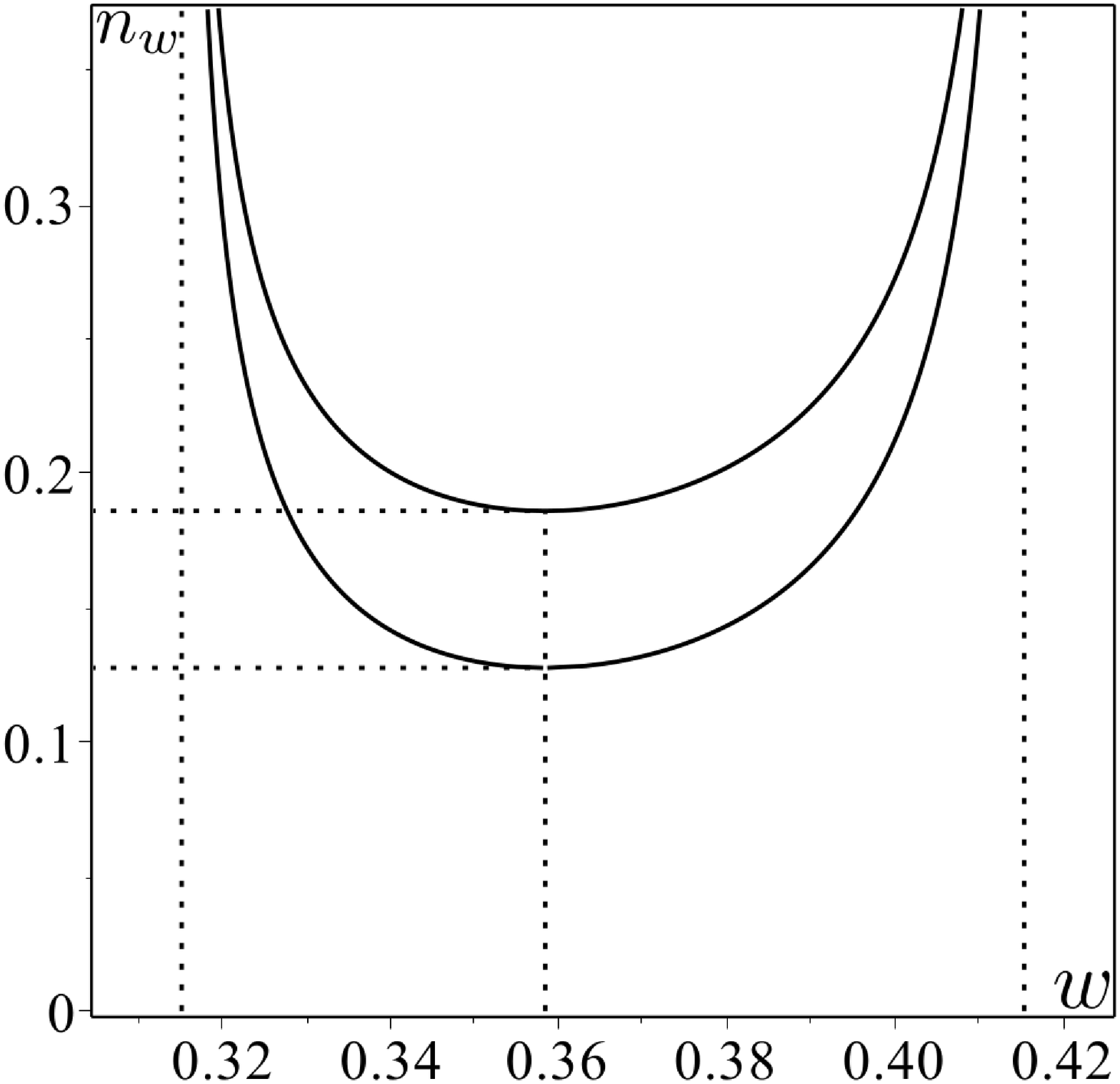}
\caption{Spectral function for ISCO, $b=2.251$, at $\zeta_{e}=5/6$. The inclination angle is $\theta_o=30^{\circ}$.
The angular velocity of the emitter is $\Omega=0.162$ and its specific energy is ${\cal E}=0.465$. The spectrum has peaks at $w_{-}= 0.315$ ($\varphi_{m}= 104^{o}7'$) and at $w_{+}=0.415$ (at $-\varphi_{m}$).
The minimal values ($0.128$ and $0.186$) of $n_{w}$ for two spectral branches are at $w_{0}=0.358$. The width parameter is $\Delta=0.274$ and the asymmetry parameter is $\delta=0.137$. One also has $N_o=0.050$.
}\label{ISCO5630}
\end{center}
\end{figure}

\begin{figure}[htb]
\begin{center}
\includegraphics[width=5cm]{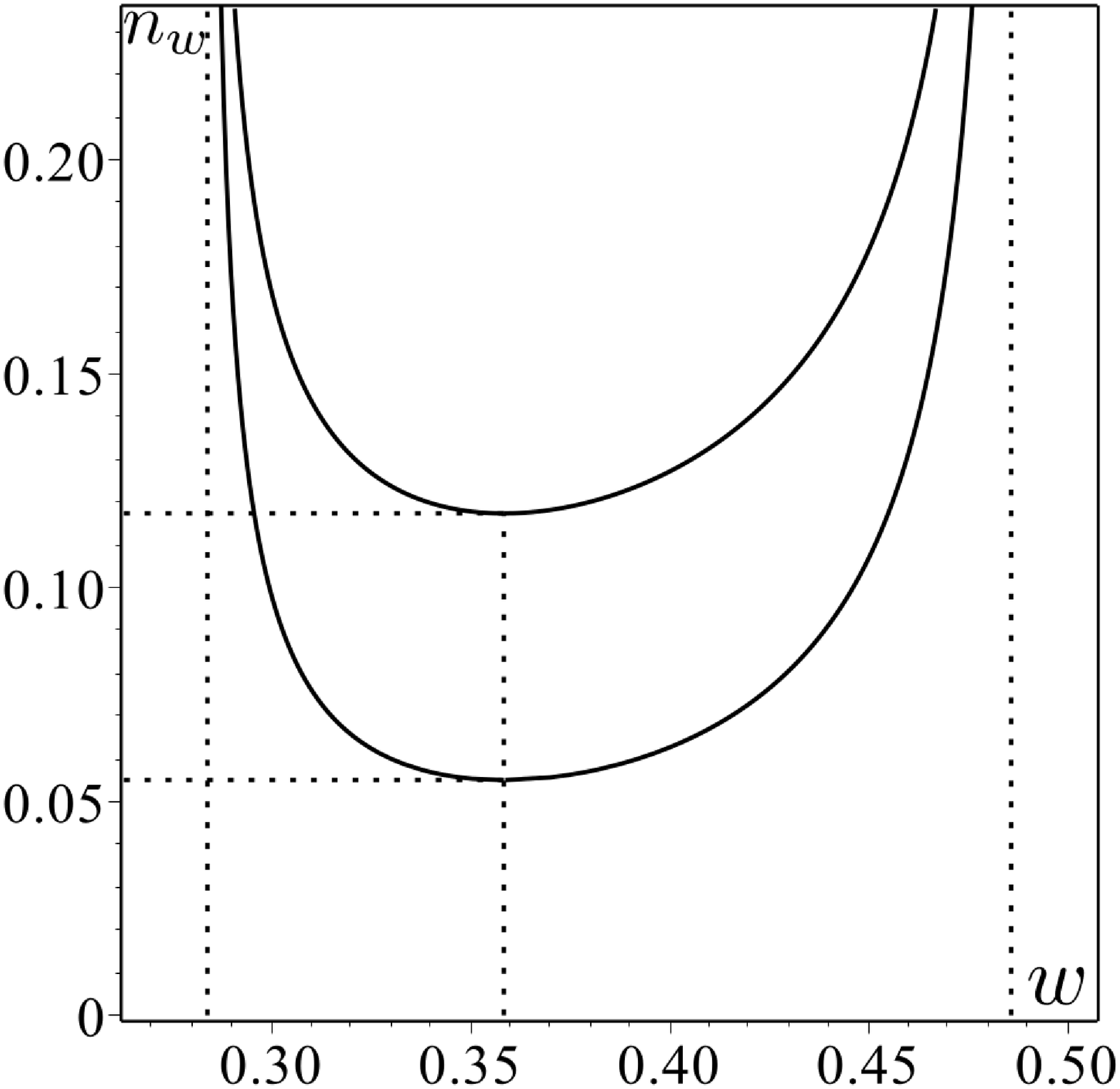}
\caption{Spectral function for ISCO, $b=2.251$, at $\zeta_{e}=5/6$. The inclination angle is $\theta_o=60^{\circ}$.
The angular velocity of the emitter is $\Omega=0.162$ and its specific energy is ${\cal E}=0.465$. The spectrum has peaks at $w_{-}= 0.284$ ($\varphi_{m}= 124^{o}6'$) and at $w_{+}=0.486$ (at $-\varphi_{m}$).
The minimal values ($0.055$ and $0.117$) of $n_{w}$ for two spectral branches are at $w_{0}=0.358$. The width parameter is $\Delta=0.525$ and the asymmetry parameter is $\delta=0.262$. One also has $N_o= 0.058$.}\label{ISCO5660}
\end{center}
\end{figure}

\begin{figure}[htb]
\begin{center}
\includegraphics[width=5cm]{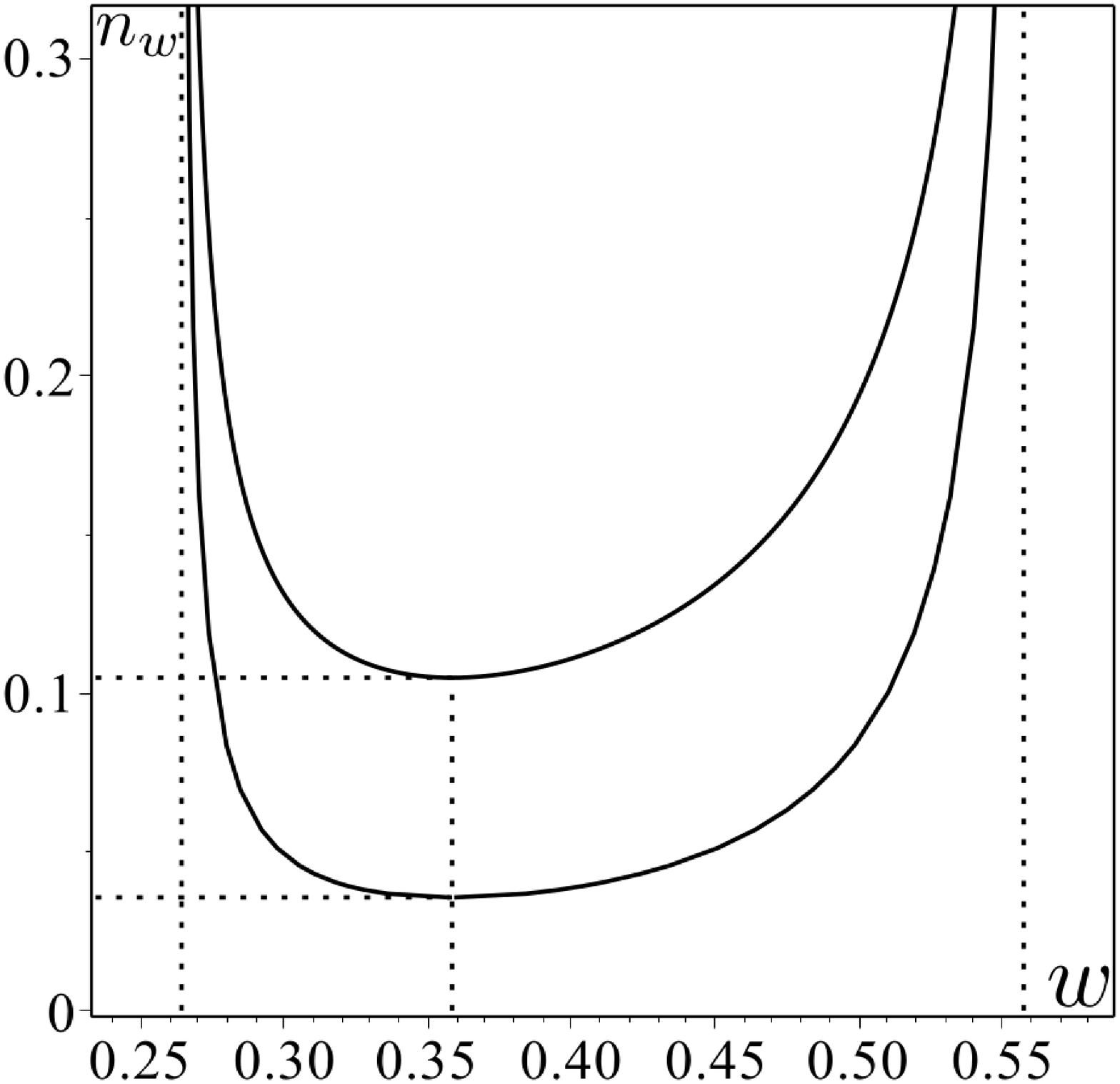}
\caption{Spectral function for ISCO, $b=2.251$, at $\zeta_{e}=5/6$. The inclination angle is $\theta_o=85^{\circ}$.
The angular velocity of the emitter is $\Omega=0.162$ and its specific energy is ${\cal E}=0.465$. The spectrum has peaks at $w_{-}= 0.264$ ($\varphi_{m}= 158^{o}6'$) and at $w_{+}=0.558$ (at $-\varphi_{m}$).
The minimal values ($0.036$ and $0.105$) of $n_{w}$ for two spectral branches are at $w_{0}=0.358$. The width parameter is $\Delta=0.714$ and the asymmetry parameter is $\delta=0.357$. One also has $N_o=0.085$.}\label{ISCO5685}
\end{center}
\end{figure}

\begin{figure}[htb]
\begin{center}
\includegraphics[width=5cm]{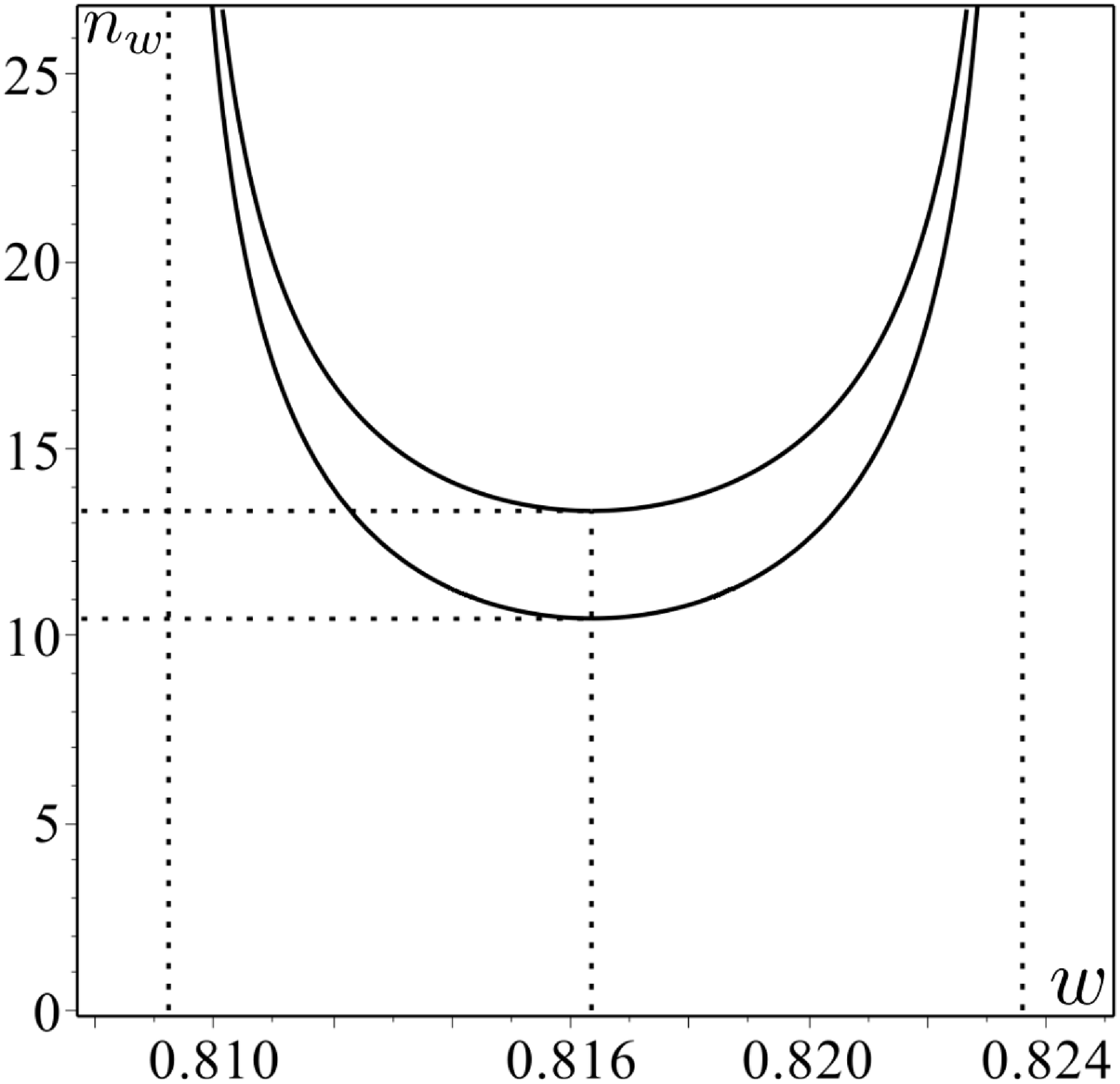}
\caption{Spectral function for SCO, $b=2.251$, at $\zeta_{e}=1/3$. The inclination angle is $\theta_o=30^{\circ}$.
The angular velocity of the emitter is $\Omega=0.005$ and its specific energy is ${\cal E}=0.817$. The spectrum has peaks at $w_{-}=0.809$ ($\varphi_{m}= 97^{o}85'$) and at $w_{+}=0.823$ (at $-\varphi_{m}$).
The minimal values ($10.46$ and $13.35$) of $n_{w}$ for two spectral branches are at $w_{0}=0.816$. The width parameter is $\Delta=0.018$ and the asymmetry parameter is $\delta=0.009$. One also has $N_o=0.390$.}\label{SCO1330}
\end{center}
\end{figure}

\begin{figure}[htb]
\begin{center}
\includegraphics[width=5cm]{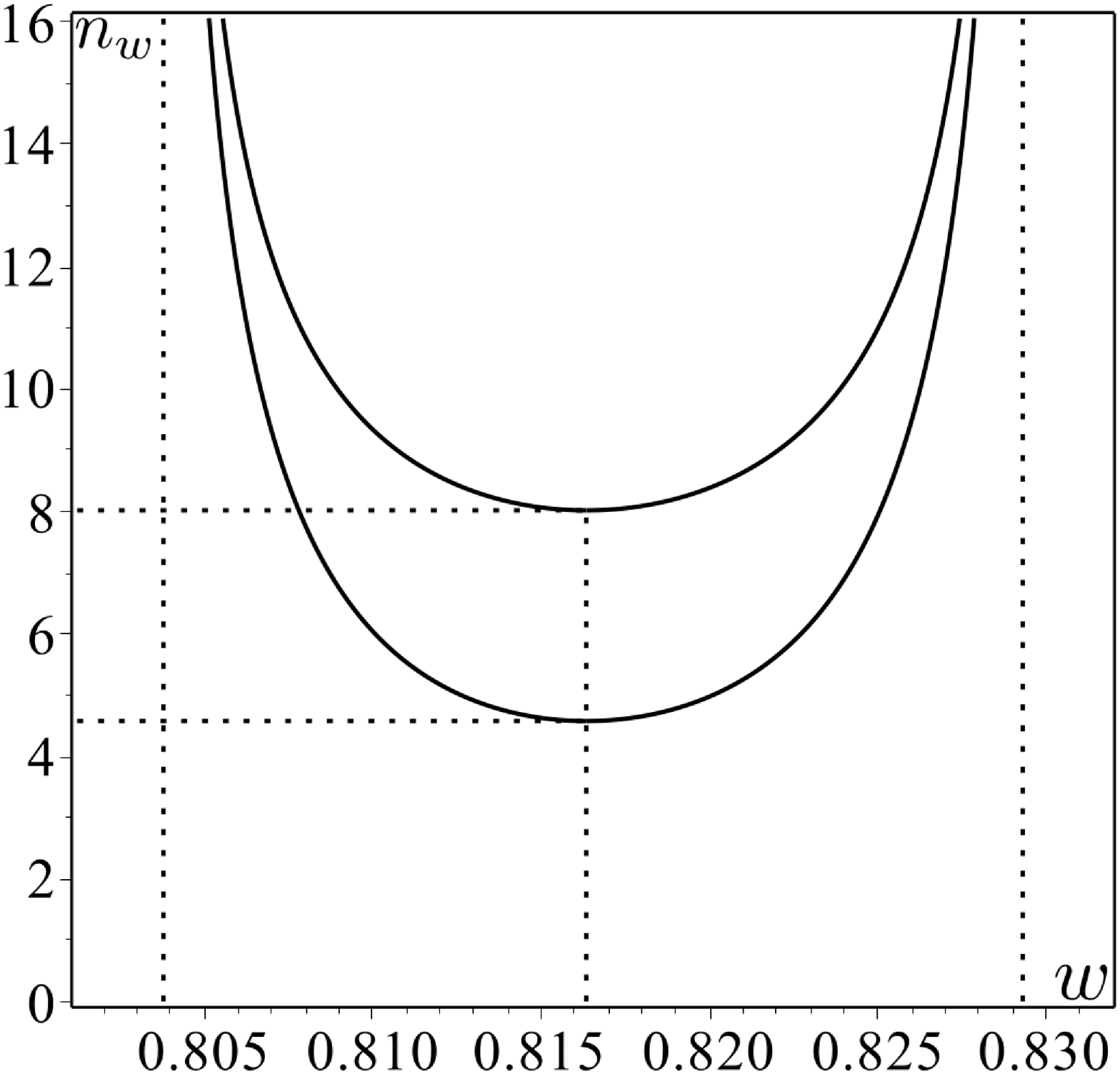}
\caption{Spectral function for SCO, $b=2.251$, at $\zeta_{e}=1/3$. The inclination angle is $\theta_o=60^{\circ}$.
The angular velocity of the emitter is $\Omega=0.005$ and its specific energy is ${\cal E}=0.817$. The spectrum has peaks at $w_{-}= 0.804$ ($\varphi_{m}= 107^{o}9'$) and at $w_{+}=0.829$ (at $-\varphi_{m}$).
The minimal values ($4.578$ and $8.018$) of $n_{w}$ for two spectral branches are at $w_{0}=0.816$. The width parameter is $\Delta= 0.031$ and the asymmetry parameter is $\delta=0.016$. One also has $N_o=0.382$.}\label{SCO160}
\end{center}
\end{figure}

\begin{figure}[htb]
\begin{center}
\includegraphics[width=5cm]{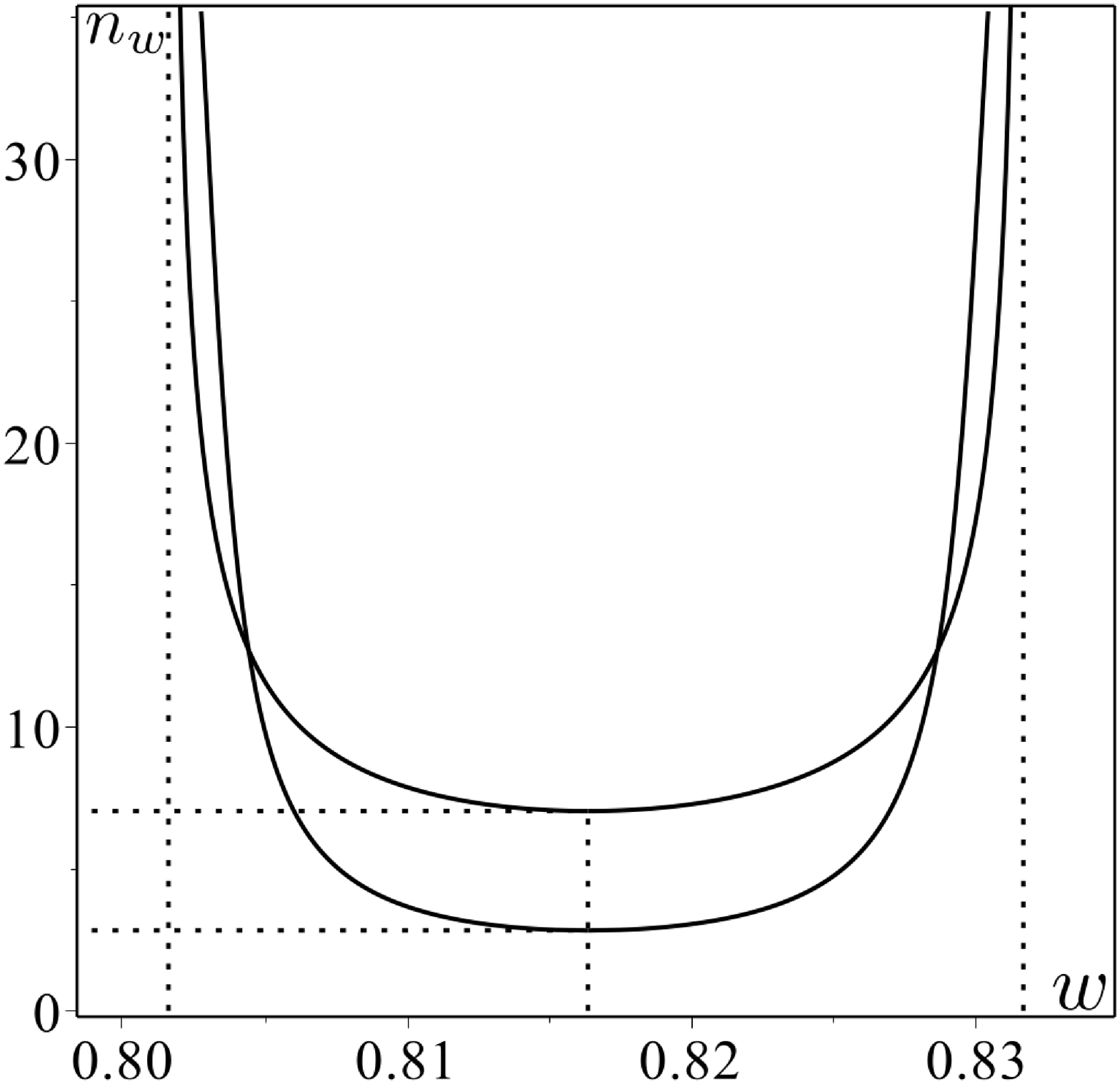}
\caption{Spectral function for SCO, $b=2.251$, at $\zeta_{e}=1/3$. The inclination angle is $\theta_o=85^{\circ}$.
The angular velocity of the emitter is $\Omega=0.005$ and its specific energy is ${\cal E}=0.817$. The spectrum has peaks at $w_{-}=0.802$ ($\varphi_{m}= 118^{o}4'$) and at $w_{+}=0.832$ (at $-\varphi_{m}$).
The minimal values ($2.836$ and $7.051$) of $n_{w}$ for two spectral branches are at $w_{0}=0.816$. The width parameter is $\Delta=0.037$ and the asymmetry parameter is $\delta= 0.019$. One also has $N_o=0.388$.
}\label{SCO1385}
\end{center}
\end{figure}

After these general remarks we present the spectral function's plots. To illustrate important features of the spectral functions we present here results for three different types of the circular motion of the emitter: (1) $\rho_e=3$, $b=0$; (2) $\rho_e=3$, $b=2.251$; and (3) $\rho_e=1.2$, $b=2.251$. In the first case the orbit is ISCO in the absence of the magnetic field. In the second one it is a SCO with the same radius $\rho_e=3$ as in (1) but in the presence of the magnetic field $b$. The last case is ISCO for the same value as in (2) of the magnetic field. These choices of the emitter's orbit parameters allow one to demonstrate the dependence of the spectral functions on the magnetic field. For each of these cases we constructed three different plots corresponding to three different values, $30^{\circ}$, $60^{\circ}$ and $85^{\circ}$ of the inclination angle $\theta_o$. These plots allow one to discuss the dependence of the spectral functions on the angular position of the distant observer with respect to the emitter's orbit. Figures~\ref{ISCO1330}--\ref{ISCO1385} present spectral functions for the case (1) and three chosen inclination angles.
Figures~\ref{ISCO5630}--\ref{ISCO5685} present similar spectral functions for the case (2) and three chosen inclination angles. The spectral functions for the case (3) and three inclination angles are shown in Figures~\ref{SCO1330}--\ref{SCO1385}.
By comparing the figures for the same inclination angle one can see that if one increases the magnetic field keeping the other parameters ($\rho_e$ and $\theta_o$) fixed, then the spectral profiles get narrow. One can see this from the values of the parameter $\Delta$ [see \eq{width}]. Namely, for the ISCO at $\rho_e=3$, $b=0$, and the inclination angle values $30^{\circ}$, $60^{\circ}$ and $85^{\circ}$ we have $\Delta=0.476,\,0.845,\,0.995$, respectively, while for the SCO at $\rho_e=3$, $b=2.251$, we have $\Delta=0.018,\,0.031,\,0.037$. This narrowing is accompanied by a general redshift of the spectral function.

To summarize, the common features of the spectrum plots are: (1) the existence of the two sharp peaks at the frequencies $w_{\pm}$; (2) the existence of two branches of the spectrum;  (3) the increase of the average redshift of the spectral frequencies for ISCO with the increase of the magnetic field; (4) the narrowing of the frequency bands with the increase of the magnetic field; (5) the asymmetry of the spectrum.

The above discussion gives simple qualitative explanations of the properties (1)-(3).
Let us briefly discuss the last two properties. The larger value of the magnetic field, the closer to the horizon is the corresponding ISCO and the greater is the redshift. Numerical calculations confirm also that the width \eq{width} decreases with the increase of $b$ (property (4)). The asymmetry (5) of the spectrum is a generic property of the broadening of the sharp spectral lines for the emitters moving near black holes. It is a result of the relativistic (Doppler) beaming effect. The calculations show that the asymmetry effect becomes more profound when the inclination angle becomes larger. 
The assymmetry parameter, as well as the width parameter for the spectral functions presented in the Figures~\ref{ISCO1330}--\ref{SCO1385} can be found in the corresponding captures.

\section{Discussion}

In this paper we discuss the radiation from emitters revolving around a magnetized non-rotating black hole. Charged particles orbits near such black holes are strongly affected by the magnetic field when the dimensionless field parameter $b$  becomes of the order of one or greater. The effect of the magnetic field depends on the direction of motion of the particle. For anti-Larmor orbits the Lorentz force is directed outwards from the black hole. The ISCO radius for the anti-Larmor particles can be close to the event horizon. Such particles on the circular orbits, after passing the ISCO limit for neutral particles at $6M$, continue their motion at SCO until they reach the critical (ISCO) radius corresponding to the given value of $b$. During this process their specific energy ${\cal E}$ decreases, so that in such process they loose slowly their energy and angular momentum  (for example, as a result of the synchrotron radiation). The maximal energy release in this process reaches 100\% in the limit $b\to\infty$.

The behavior of the Larmor particles is quite different. For $b>0$ their circular orbits can also have radius less than $6M$. However, in order to move on such orbits they need to receive some additional energy. This means, that one can expect that either such particles are accumulated during some period of time near $6M$ orbits, or they  simply fall down directly to the black hole similarly to neutral particles. This might have quite interesting consequence: the effect of the spatial separation of charge. We do not discuss this effect here. In this paper we focused to the radiation emitted by anti-Larmor particles moving close to the black hole in the presence of the magnetic field. Namely, we analysed two problems: (1) Images of such orbits, and (2) Spectral broadening of the emission received from a moving emitted by a distant observer. Both the problems require ray-tracing of photons in the Schwarzschild metric, the problem which is well known and discussed in details in literature. However, we apply this ray-tracing to orbits which are closer than $6M$ to the back hole horizon. Namely, these orbits
are interesting  for magnetized black holes. Similar remarks can be done for the spectral broadening problem. In magnetized black holes both the position of the circular orbits and angular velocity of the evolution are different from the Keplerian case, which was studied earlier.

Images of the anti-Larmor orbits close to the horizon of the magnetized black holes are presented in Figure~\ref{F11}. The main conclusion is that in the limit of large magnetic field $b$ the ISCO image structure is simplified. In this limit the image basically consists of two parts: (1) a semicircle inside the shadow domain region, and (2) practically straight line in the equatorial plane. The first part of the image is generated by rays from the part of the orbit `behind' the black hole, while the latter one is form by direct rays emitted `in front' of the black hole.

We discussed and compared in the previous section the spectral functions for sharp line broadening. The $\delta$-function-like radiation spectrum for the monochromatic radiation of the charged anti-Larmor emitter is registered by a distant observer as a broadened spectrum. This is the result of two effects: Doppler effect and gravitational redshift. The width of the spectral function is determined by the periodic Doppler blue- and red-shift. It is proportional to the angular velocity $\Omega$ of the emitter and it vanishes in the limit $b\to \infty$.
The closer the orbit of the emitter is to the horizon, the larger is influence of the gravitational field on the spectrum. This effect results in the total redshift of the spectral frequencies.  One can summarize the generic properties of the spectral broadening for magnetized black hole as follows: when the magnetic field $b$ increases, both the width of the spectrum and its average frequency decrease.

It would be interesting to compare the spectral broadening in magnetized black holes with a similar effect in rotating black holes. Action of the dragging effect of the black hole on neutral particles is similar to the effect of the magnetic field on charged particles: in the both cases ISCO's for (1) Co-rotating particles in the Kerr metric, and (2) anti-Larmor particles in the magnetic field can be located arbitrarily close to the horizon. However, there is a big difference between these two cases. A particle close to the Kerr black hole is co-rotating with the black hole, so that its angular velocity tends to the black hole angular velocity. In the magnetized black holes for anti-Larmor orbits close to the horizon the angular velocity tends to zero (we discussed the reason of this behavior in subsection D of Sec. II). Hence, one can expect that the width of the spectra for the radiation emitted by anti-Larmor particles moving close to the horizon of magnetized black holes must be smaller than the corresponding width for neutral particles in rotating black holes. Anyway, it would be interesting to perform the calculations of the spectral function broadening for a general case of a magnetized rotating black hole.

In the present paper we made two simplifying assumptions. We used special anzats for  the form of the magnetic field. In realistic black holes one cannot expect that the magnetic field is homogeneous and extends to infinity. However, for the motion of a charged particle in the equatorial plane and in the black hole vicinity this approximation might be reasonable. It is easy to extend the results for other types of a regular magnetic field, e.g., for a dipolar magnetic field around a static black hole (see, e.g., \cite{PrVa}). Moreover, a model of the homogeneous magnetic field is a good approximation for more realistic magnetic fields generated by currents in a conducting accretion disk, provided the size of the black hole is much smaller than the size of the disk (see, e.g., discussion in \cite{Pet:74}).

Another assumption was that the radiating particles are localized at the infinitely thin ring of a fixed radius. In reality, one can expect that there exist some distribution of the anti-Larmor emitters extended from $6M$ to their ISCO radii. In order to obtain the spectrum of the emission from such a ring of finite size, one needs to perform an additional integration of the obtained spectra with some weight function, which describes the distribution of the emitters within this ring. Such an averaging would smear infinite peaks and made them finite.

In spite of the made assumptions the main conclusions of the present work seem to be quite robust. Namely, for anti-Larmor emitters in magnetized black holes ISCO's are close to the horizon and dominant effects for this inner domain of radiation result in bigger redshift and narrowing of the spectrum. This allows one to hope that observations of the broadening in the Iron K$\alpha$ lines in magnetized black holes can provide us with the direct information about the magnetic field in the black hole vicinity.

\acknowledgments

We thank Ted Jacobson for his remarks and suggestions proposed during the Peyresq 18 meeting, that stimulated our work on the problem of the spectral line broadening in magnetized black holes. The authors are grateful to the Natural Sciences and Engineering Research Council of Canada for its support. One of the authors (V.F.) thanks the Killam Trust for its financial support.

\appendix

\section{$\ell-$parametrization of the orbit}

For light emitted  by the emitter on the circular orbit of the inverse radius $\zeta_e$, that reaches a distant observer, there exist a relation between its angular momentum $\ell$, and  the angle $\varphi$ where it was emitted. This relation follows from \eq{BBF}. In the numerical calculations in Sec.~VI it is convenient to consider $\varphi$ as a function of $\ell$. Let us discuss the properties of this function.

From \eq{BBF} one finds
\be\n{dP}
{d\Phi\over d\varphi} {d\varphi\over d\ell}={dB_{\pm}(\ell,\zeta_e)\over d\ell}\, .
\ee
Using the definition of $\Phi$ in \eq{BBF} one obtains
\be\n{FFFF}
{d\Phi\over d\varphi}={\sin\varphi \sin\theta_o\over \sqrt{\sin^2\varphi+\cos^2\theta_o \cos^2\varphi}}\, .
\ee
Hence, $d\Phi/d\varphi$ is positive for $\varphi\in (0,\pi)$ and negative for $\varphi\in (-\pi,0)$. Using the definition \eq{BBB} of the function $B(\ell,\zeta_e)$ one gets
\be
{dB(\ell,\zeta_e)\over d\ell}=\int_0^{\zeta_e}{d\zeta\over (1-\ell^2(1-\zeta)\zeta^2)^{3/2}}\, .
\ee
From the above relations one concludes that for direct rays the sign of $d\varphi/d\ell$ coincides with the sign of $\varphi$, that is it is positive for $\varphi\in(0,\pi)$ and negative for the other segment of the trajectory.

Let us discuss now the case of indirect rays. Let us recall that a null ray has a radial turning point only when $\ell>\ell_*=3\sqrt{3}/2$. The inverse radius of this turning point $\zeta_m$ is a solution of the equation
\be\n{zzll}
(1-\zeta_m)\zeta_m^2=\ell^{-2}\, ,
\ee
and it belongs to the interval $(0, 2/3)$. Differentiating this relation with respect to $\ell$ one gets
\be
{d\zeta_m\over d\ell}=-{2\over \ell^3 \zeta_m(2-3\zeta_m)}<0\, .
\ee

Using the definition \eq{CC} of the function $C(z)$ one finds
\be
{dC(z)\over dz}=\int_0^1{ (y^4-3y^2+3) dy\over Z^{3/2}}\, .
\ee
Let us notice that both the derivatives, $dB/d\ell$ and $dC/dz$, are positive definite. (For $dC/dz$ this is because $y^4-3y^2+3>0$).

Using \eq{BBF} we obtain
\be\n{dBB}
{d\Phi\over d\varphi} {d\varphi\over d\ell}={dB_{-}(\ell,\zeta)\over d\ell}=
2 {dC(z_m)\over d\zeta_m}{d\zeta_m\over d\ell}-{dB(\ell,\zeta_e)\over d\ell} \, .
\ee
Both of the terms in the right-hand side of \eq{dBB} are negative. \eq{FFFF} implies that the sign of $d\Phi/d\varphi$ coincides with the sign of $\varphi$. To summarize, $d\varphi/d\ell$ is negative in the interval $\phi\in (0,\pi)$ and positive in the other half of the circle.

\section{Radiation from a point-like source}

In this appendix we discuss the following problem. Suppose one has a point-like source emitting photons. Denote its 4-velocity by $\BM{u}$. Denote by $\tau$ proper time along the emitters world line. Choose a moment of time $\tau_e$  and consider a local frame $\BM{e}_{a}$ $(a=0,\ldots,3)$ at this point. We choose $\BM{e}_0=\BM{u}$. We consider one of the emitted at $\tau_e$ photons and call it a {\em reference photon}. Its 4-momentum be $\BM{p}$ can be written as follows
\be
\BM{p}=\omega_e (\BM{u}+\BM{N})\hh \omega_e=-(\BM{p},\BM{u})\, .
\ee
$\omega_e$ is the frequency of the photon in the rest frame of the emitter and $\BM{N}$ is a unit vector orthogonal to $\BM{u}$. It determines the spatial direction of the reference photon. We choose the vector $\BM{e}_1$ to coincide with $\BM{N}$.
The other two unit vectors $\BM{e}_2$ and $\BM{e}_3$ of the frame are chosen to be orthogonal to both the vectors $\BM{e}_0$ and $\BM{e}_1$. They are fixed up to a rotation and uniquely (up to the orientation) determine a 2-plane $\Pi$, orthogonal to $\BM{e}_0$ and $\BM{e}_1$. We denote the corresponding bi-vector by $\BM{e}_2\wedge\BM{e}_3$.

Consider a bundle of photons emitted in the direction close to $\BM{N}$ within a solid angle $\Delta\Omega_e$ which is determined by two vectors $\Delta_1\BM{N}$ and $\Delta_2\BM{N}$, orthogonal to $\BM{N}$. The vectors of this bundle can be parameterized as follows
\be
\BM{N}+\alpha_1 \Delta_1\BM{N}+\alpha_2 \Delta_2\BM{N}\hh\alpha_{1,2}\in(-1/2,1/2)\, .
\ee
The solid angle $\Delta\Omega_e$ coincides with the area of the parallelogram in $\Pi$ determined by these two vectors, which is
\be
\Delta\Omega_e=\|\Delta_1\BM{N}\wedge\Delta_2\BM{N}\|=|\Delta_1 N^2 \Delta_2 N^3-\Delta_1 N^3 \Delta_2 N^2|\, ,
\ee
where $\Delta_1 N^i$ and $\Delta_2 N^i$ are components of $\Delta_1\BM{N}$ and $\Delta_2\BM{N}$ in the 2D basis $\{\BM{e}_2,\BM{e}_3\}$.

We assume now that the radiation of the emitter is isotropic and denote by ${\cal N}\Delta\tau_e$ total number of photons emitted during the time interval $\Delta\tau_e$ of the proper time in the frame co-moving with the emitter. Then the corresponding number of photons emitted in the solid angle $\Delta\Omega_e$ is
\be\n{NE}
n_e\Delta\tau_e={{\cal N}\over 4\pi} \Delta\Omega_e\Delta\tau_e\, .
\ee

The solid angle $\Delta\Omega_e$ can also be determined by the relation
\be\n{www}
\BM{u}\wedge \BM{N}\wedge \Delta_1\BM{N} \wedge \Delta_2\BM{N}=\pm \Delta\Omega_e \BM{E}\, ,
\ee
where $\BM{E}$ is a unit 4-form
\be
\BM{E}=\BM{e}_0\wedge \BM{e}_1\wedge \BM{e}_2\wedge \BM{e}_3\, .
\ee
Adding to $\BM{N}$ the vector $\BM{u}$ does not change the value of the wedge product in the left-hand side of \eq{www}. Similarly, adding the vectors proportional to $\BM{u}$ and $\BM{N}$ to the vectors $\Delta_1\BM{N}$ and $\Delta_2\BM{N}$ does not change this wedge product. As a result one can rewrite \eq{www} in the form
\be\n{EEEE}
{1\over \omega_e^3}\BM{u}\wedge \BM{p}\wedge \Delta_1\BM{p} \wedge \Delta_2\BM{p}=\pm \Delta\Omega_e \BM{E}\, ,
\ee
or, that is equivalent, in the form
\be\n{PPPP}
\Delta\Omega_e=\pm {1\over \omega_e^3} e_{\mu_1 \mu_2 \mu_3 \mu_4} u^{\mu_1} p^{\mu_2} \Delta_1 p^{\mu_3}\Delta_2 p^{\mu_4}\, .
\ee
Here $e_{\mu_1 \mu_2 \mu_3 \mu_4}=\sqrt{-g}\epsilon_{\mu_1 \mu_2 \mu_3 \mu_4}$ is totally skew symmetric tensor and  $\epsilon_{\mu_1 \mu_2 \mu_3 \mu_4}$ is the Levi-Civita symbol.


\begin{thebibliography}{99}

\bi{Na:05}\NJP{R. Narayan, }{7}{199}{2005}
\bi{ReBe:77} \APJ{C.S. Reynolds and M.C. Begelman, }{488}{109}{1997}
\bi{BrChMi} \APJ{B.C. Bromley, L. Chen and W.A. Miller, }{475}{57}{1997}
\bi{ZaNuPaIn:05} A.F.Zakharov, A.A.Nucita, F.DePaolis and G.Ingrosso, talk at the
Workshop on High Energy Physics and Field Theory (Protvino, Russia, 2004);
e-print arXiv:gr-qc/0507118 (2005).
\bi{Za:07} A.F.Zakharov, Physics of Atomic Nuclei, {\bf 70}, 159 (2007).
\bi{KaMi:04} N. Kawanaka and S. Mineshige, in {\em Proceedings of 22nd Texas Simposium on Relativistic Astrophysics}, 2105 (2004).
\bibitem{LtoN95} Y. Tanaka et al., \emph{Letters to Nature} {\bf 375}, 659 (1995).
\bibitem{CuBa:73} C. T. Cunningham, J. M. Bardeen, \emph{Astrophys. J.} {\bf 183},  237 (1973).
\bibitem{Cu:75} C. T. Cunningham, \emph{Astrophysical Journal} {\bf 202}, 788 (1975).
\bibitem{CaFaCa:98} A. Cadez, C. Fanton and M. Calvani, \emph{New Astronomy} {\bf 3}, 647 (1998).
\bibitem{FrKlNe:00} S. Frittelli, T. P. Kling and E. T. Newman, \emph{Phys. Rev. D} {\bf 61}, 064021 (2000).
\bibitem{FuWu:04} S. V. Fuerst, K. Wu, \emph{Astron. Astrophys.}  {\bf 424}, 733 (2004).
\bi{FaIwRe:00} A.C. Fabian, K. Iwasawa, C.S. Reynolds and
A.J. Young, Publications of the Astronomical Society
of the Pacific (PASP), {\bf 112}, 1145 ( 2000)
\bi{ReNo:03} \PRep{C.S. Reynolds and M.A. Nowak, }{377}{389}{2003}
\bi{FuTs:04} \APJ{K. Fukumura and S. Tsuruta, }{613}{700}{2004}
\bi{Jo:12} P. Jovanovi\'{c}, New Astronomy Reviews, {\bf 56}, issues 2-3, 33 (2012).
\bi{BaHa:98}\RMP{S.A. Balbus and J.F. Hawley, }{70}{1}{1998}
\bi{Kr:01} J.H. Krolik, in {\em Proceedings of 20nd Texas Simposium on Relativistic Astrophysics}, AIP Conf. Proc. {\bf 586}, 674 (2001).
\bi{Ea:13}\NA{R.P. Eatough et al., }{501}{391}{2013}
\bi{Fa}\ASA{H. Falke and S. Markoff, }{362}{113}{2000}
\bi{Mo}\APJ{M. Moscibrodzka, C.F. Gammie, J.C. Dolence, H. Shiokawa and P.K. Leung, }, {706}{497}{2009}
\bi{De}\APJ{J. Dexter, E. Agol, P.C. Fragile and J.C. McKinney, }{717}{1092}{2010}
\bi{Zn:76}\NA{R. Znajek, }{262}{270}{1976}
\bi{BlZn}\MN{R.D. Blandford and R. Znajek, }{179}{433}{1977}
\bi{MEM}\BOOK{K.S. Thorne, R.H. Price and D.A. MacDonald}{Black Holes: The Membrane Paradigm}{Yale University Press}{1986}
\bi{KoShKuMe:02}\SC{S. Koide, K. Shibata, T. Kudoh and D.L. Meier, }{295}{1688}{2002}
\bi{Ko:04}\APJ{S. Koide, }{606}{L45}{2004}
\bi{Si:09}\ASA{N. A. Silant'ev, M. Yu. Piotrovich, Yu. N. Gnedin, and T. M. Natsvlishvili, }, {507}{171}{2009}
\bi{Pi:00}\ASL{M. Yu. Piotrovich, N. A. Silant'ev, Yu. N. Gnedin, }, {36}{389}{2010}
\bi{SiGn:13}\AB{N. A. Silant'ev, Yu. N. Gnedin, S.D. Buliga, M. Yu. Piotrovich, and T. M. Natsvlishvili, }, {68}{14}{2013}
\bi{Za:03}\MN{A.F. Zakharov, N.S. Kardashev, V.N. Lukash and S.V. Repin, }{342}{1325}{2003}
\bibitem{AG} A. N. Aliev and D. V. Gal'tsov, Sov. Phys. Usp. {\bf 32}, 75 (1989).
\bi{FS} V. P. Frolov and A. A. Shoom, Phys. Rev. D {\bf 82}, 084034 (2010).
\bibitem{MTW} C.~W.~Misner, K.~S.~Thorne and J.~A.~Wheeler, {\em Gravitation} (W. H. Freeman and Co., San Francisco, 1973).
\bi{FF} \PRD{V.P. Frolov, }{85}{024020}{2012}
\bibitem{Wald} R. M. Wald, Phys. Rev. D {\bf 10}, 1680 (1974).
\bibitem{FZ} V.P. Frolov and A. Zelnikov, {\em Introduction to Black Hole Physics}, Oxford University Press, 2011.
\bibitem{PrVa} A. R. Prasanna and R. K. Varma, Pramana {\bf 8}, 229 (1977).
\bibitem{Pet:74} J. A. Petterson, Phys. Rev. D {\bf 10}, 3166 (1974).




\end{thebibliography}
\end{document}